\documentclass[11pt,a4paper]{article}

\pdfoutput=1
\usepackage{amsmath,amsfonts,amsbsy,amssymb,enumerate,array}
\usepackage{graphicx}
\usepackage[usenames]{color}
\usepackage[english]{babel}
\usepackage{fancybox}
\usepackage{slashed}
\usepackage{chngpage} % allows for temporary adjustment of side margins

\definecolor{AV}{rgb}{0.65,0.0,0}
\definecolor{DT}{rgb}{0,0,0.65}

\newcommand{\captionfonts}{\small}

\makeatletter  % Allow the use of @ in command names
\long\def\@makecaption#1#2{%
  \vskip\abovecaptionskip
  \sbox\@tempboxa{{\captionfonts #1: #2}}%
 \ifdim \wd\@tempboxa >\hsize
    {\captionfonts #1: #2\par}
  \else
    \hbox to\hsize{\hfilsection\box\@tempboxa\hfil}%
  \fi
  \vskip\belowcaptionskip}
\makeatother   % Cancel the effect of \makeatletter

\newcommand{\nn}{\nonumber}
\newcommand{\rom}[1]{\mathrm{#1}}
\definecolor{purple}{rgb}{0.7,0,1}
\definecolor{grey}{rgb}{0.5,0.5,0.5}

\newcommand{\expe}[1]{\textrm{e}^{#1}}
\newcommand{\fr}{\frac}
\newcommand{\f}{\frac}

\newcommand\half{\frac{1}{2}}

\newcommand \RR{{\mathbb{R}}}

\newcommand\cA{\mathcal{A}}
\newcommand\cB{\mathcal{B}}
\newcommand\cC{\mathcal{C}}
\newcommand\cD{\mathcal{D}}
\newcommand\cE{\mathcal{E}}
\newcommand\cF{\mathcal{F}}
\newcommand\cG{\mathcal{G}}
\newcommand\cH{\mathcal{H}}
\newcommand\cI{\mathcal{I}}
\newcommand\cJ{\mathcal{J}}
\newcommand\cK{\mathcal{K}}
\newcommand\cL{\mathcal{L}}
\newcommand\cN{\mathcal{N}}
\newcommand\cO{\mathcal{O}}
\newcommand\cP{\mathcal{P}}
\newcommand\cQ{\mathcal{Q}}
\newcommand\cR{\mathcal{R}}
\newcommand\cS{\mathcal{S}}
\newcommand\cT{\mathcal{T}}
\newcommand\cU{\mathcal{U}}
\newcommand\cW{\mathcal{W}}
\newcommand\cX{\mathcal{X}}
\newcommand\cY{\mathcal{Y}}

\newcommand{\mH}{\mathcal{H}}

\newcommand{\mL}{\mathcal{L}}

\newcommand{\mN}{\mathcal{N}}

\newcommand{\mR}{\mathcal{R}}

\newcommand{\mV}{\mathcal{V}}
\newcommand{\mZ}{\mathcal{Z}}

\def\La{\Lambda}
\newcommand{\ze}{\zeta}

\def\g{\mathfrak{g}}
\def\so{\mathfrak{so}}

\newcommand{\si}{\sigma}         \newcommand{\Si}{\Sigma}

\newcommand{\ch}{\chi}
\newcommand{\om}{\omega}        

\newcommand{\ti}{\tilde}
\newcommand{\ov}{\overline}
\newcommand{\sq}{\sqrt}
\def\beq{\begin{eqnarray}}
\def\eeq{\end{eqnarray}}

\def \RR{{\mathbb{R}}}

\usepackage[colorlinks=true,      linkcolor=blue,      urlcolor=blue,      filecolor=blue,      citecolor=blue,
      pdfstartview=FitH,     pdfpagemode=UseNone,      bookmarksopen=true]{hyperref}  % LINKS

\usepackage[all]{hypcap}     %should be loaded AFTER hyperref, which should otherwise be loaded last.

\begin{document}

\numberwithin{equation}{section}

%%%%%%%%%%%%%%%%%%%%%%%%%%%%%%%%%%%%%%%%%%%%%%%%%%%%%%%%%%%%%%%%%%%%%%%%%%%%%
%                       DEFINITIONS

\newcommand{\be}{\begin{equation}}
\newcommand{\ee}{\end{equation}}
\newcommand{\bea}{\begin{eqnarray}\displaystyle}
\newcommand{\eea}{\end{eqnarray}}

\def\eq#1{(\ref{#1})}

%%%%%%%%%%%%%%%%%%%%%%%%%%%%%%%%%%%%%%%%%%%%%%%%%%%%%%%%%%%%%%%%%%%%%%%%%%%%%
%                        Greek letters

\def\a{\alpha}  \def\b{\beta}   \def\c{\chi}
\def\g{\gamma}  \def\G{\Gamma}  \def\e{\epsilon}
\def\vep{\varepsilon}   \def\tvep{\widetilde{\varepsilon}}
\def\vf{\varphi}  \def\m{\mu}  \def\mub{\ov \mu}
\def\n{\nu}  \def\nub{\ov \nu}  \def\o{\omega}
\def\O{\Omega}  \def\r{\rho}  \def\k{\kappa}
\def\kab{\ov \kappa}  \def\s{\sigma}
\def\t{\tau}  \def\th{\theta}  \def\sb{\ov\sigma}  \def\S{\Sigma}
\def\l{\lambda}  \def\L{\Lambda}  
\def\p{\partial}

\newcommand{\gt}{\tilde{\gamma}}

%%%%%%%%%%%%%%%%%%%%%%%%%%%%%%%%%%%%%%%%%%%%%%%%%%%%%%%%%%%%%%%%%%%%%%%%%%%%%
%              Calligraphic & Blackboard letters etc

\def\cA{{\cal A}} \def\cB{{\cal B}} \def\cC{{\cal C}}
\def\cD{{\cal D}} \def\cE{{\cal E}} \def\cF{{\cal F}}
\def\cG{{\cal G}} \def\cH{{\cal H}} \def\cI{{\cal I}}
\def\cJ{{\cal J}} \def\cK{{\cal K}} \def\cL{{\cal L}}
\def\cM{{\cal M}} \def\cN{{\cal N}} \def\cO{{\cal O}}
\def\cP{{\cal P}} \def\cQ{{\cal Q}} \def\cR{{\cal R}}
\def\cS{{\cal S}} \def\cT{{\cal T}} \def\cU{{\cal U}}
\def\cV{{\cal V}} \def\cW{{\cal W}} \def\cX{{\cal X}}
\def\cY{{\cal Y}} \def\cZ{{\cal Z}}

\def\mC{\mathbb{C}} \def\mP{\mathbb{P}}
\def\mR{\mathbb{R}} \def\mZ{\mathbb{Z}}
\def\mT{\mathbb{T}} \def\mN{\mathbb{N}}
\def\mH{\mathbb{H}} \def\mX{\mathbb{X}}

\def\one{{\hbox{\kern+.5mm 1\kern-.8mm l}}}

\begin{flushright}
IP-BBSR-2016-5
\end{flushright}

\vspace{7mm}

\begin{center}

{\LARGE \textsc{Smooth non-extremal D1-D5-P solutions}} \\

%\vspace{6mm}

%{\huge \textsc{}} \\

\vspace{6mm}

{\LARGE \textsc{as charged gravitational instantons}}

\vspace{15mm}

{\large
\textsc{ Bidisha Chakrabarty${}^{1}$, ~ Jorge V.~Rocha${}^{2}$, ~ Amitabh Virmani${}^{1}$
}}

\vspace{13mm}

${}^{1}${Institute of Physics, Sachivalaya Marg, \\ Bhubaneshwar, India -- 751005 }\\

\vspace{4mm}

${}^{2}$
{Departament de F\'isica Fonamental, Institut de Ci\`encies del Cosmos (ICCUB), \\ Universitat de Barcelona, Mart\'i i Franqu\`es 1, E-08028 Barcelona, Spain}

\vspace{18mm}

\textsc{Abstract}

\end{center}

\begin{adjustwidth}{10mm}{10mm} % to adjust the L and R margins

\vspace{0.3 cm}

{\small
\noindent
We present an alternative and more direct construction of the non-supersymmetric D1-D5-P supergravity solutions found by Jejjala, Madden, Ross and Titchener.
We show that these solutions --- with all three charges and both rotations turned on --- can be viewed as a charged version of the Myers-Perry instanton. 
We present an inverse scattering construction of the Myers-Perry instanton metric in Euclidean five-dimensional gravity. 
The angular momentum bounds in  this construction turn out to be precisely the  ones necessary for the smooth microstate geometries. 
We add charges on the Myers-Perry instanton using appropriate SO$(4,4)$ hidden symmetry transformations. 
The full construction can be viewed as an extension and simplification of a previous work by Katsimpouri, Kleinschmidt and Virmani.
}

\end{adjustwidth}

\thispagestyle{empty}

\newpage

\baselineskip=13.5pt
\parskip=3pt

\tableofcontents

\section{Introduction}
One of the key steps that advanced the study of  three-charge supersymmetric black hole microstates was the rewriting by Giusto and Mathur \cite{Giusto:2004kj} of the first example of a smooth geometry in the fibered form, thus making the connection with the classification of supersymmetric solutions. This exercise led to the realisation that the four-dimensional base space for such solutions had to be of the so-called ``pseudo-hyper-K\"ahler'' form, which paved the way for generalisations to the multi-center solutions  \cite{Bena:2005va, Berglund:2005vb}. 

It is natural to hope that understanding the known non-extremal microstates \cite{Jejjala:2005yu, Giusto:2007tt, AlAlawi:2009qe, Banerjee:2014hza, Bena:2009qv, Bossard:2014yta, Bena:2015drs} from various possible perspectives will shed  light on how to go about constructing more general non-extremal microstates.
Drawing movitation from  properties of the supersymmetric solutions, one such study was performed in reference \cite{Gimon:2007ps} for the solutions found by Jejjala, Madden, Ross, and Titchener (JMaRT) \cite{Jejjala:2005yu}. They found that upon dimensional reduction from 6d to 5d, the 5d solution features locally non-supersymmetric orbifold singularities. Upon further reduction to 4d, they found that the two singularities are connected by a conical singularity. The presence of  the conical singularity does not allow for an unambiguous association of brane charges to the two  centers. This led the authors to conclude that the picture of ``half-BPS atoms'' making up the multiple centers of supersymmetric microstates does not extent to the non-supersymmetric ones in any easy way. One must consider more general kinds of basic building blocks.

In this paper we add a new dimension to this discussion. We show that the JMaRT solution can also be thought of as a charged version of Euclidean five-dimensional Myers-Perry instanton trivially lifted to six dimensions by the addition of a flat timelike direction. Gravitational instantons in four-dimensions have received much attention under the Euclidean Gravity paradigm, though their higher-dimensional cousins are not so well explored. For the cases where these objects have been explored, their classification is presented in terms of turning points of various degenerating Killing vectors \cite{Chen:2010zu}; more precisely in terms of the so-called rod structure \cite{Emparan:2001wk, Harmark:2004rm,Hollands:2007aj}. Since for the non-supersymmetric microstates only spacelike Killing vectors degenerate, it is natural to expect that non-supersymmetric microstates are closely related to gravitational instantons.

For the construction of the multi-center supersymmetric solutions this connection is the key element \cite{Bena:2005va, Berglund:2005vb}.  In these constructions the four-dimensional base space is taken to be multi-center Gibbons-Hawking instanton.  For non-extremal microstates such a link has also been explored, though not yet in a fully systematic way. For example, the first generalisation  \cite{Giusto:2007tt} of the JMaRT solution was constructed by adding appropriate charges to the so-called Kerr-Taub-Bolt instanton. Similar ideas, in different guises, were also used in references \cite{Bobev:2009kn, Banerjee:2014hza, Bossard:2014ola, Bossard:2014yta}. More recently, these and a related circle of ideas have led to the construction of the first example of non-extremal multi-bubble microstate geometries \cite{Bena:2015drs}.

It had been anticipated that the JMaRT solution has a close connection to gravitational instantons (see e.g. comments in \cite{Giusto:2007tt, Bossard:2014ola}), though it has never been made precise. A connection was established in reference \cite{Katsimpouri:2014ara} where it was highlighted that the JMaRT metric can be related to the Myers-Perry instanton metric via a simple analytic continuation.  In this paper we extend and simplify that construction. There are several differences: we consider both angular momentum and all three charges, whereas reference \cite{Katsimpouri:2014ara} only dealt with the case of two-charges and a single rotation. We work with the well developed Belinski-Zakharov inverse scattering method \cite{BZ, Pomeransky:2005sj}, as opposed to the Breitenlohner-Maison method \cite{Breitenlohner:1986um, Katsimpouri:2013wka, Katsimpouri:2012ky, Chakrabarty:2014ora} used in \cite{Katsimpouri:2014ara}.  Moreover, for adding charges we do reductions over the standard angular coordinates $\psi$ and $\phi$ as opposed to linear combinations of these coordinates as was done there. We use timelike reduction to go from 4d to 3d, as opposed to  \cite{Katsimpouri:2014ara} where the timelike reduction was used to go from 6d to 5d.
These points considerably simplify the calculations and make the full construction more accessible.

The rest of the paper is organised as follows. 
In section \ref{JMaRT_main} we gather our main ideas relegating all detailed calculations to the appendices. In section \ref{sec:MPins} we present the Myers-Perry instanton metric. 
In section \ref{sec:Weyl} we 
perform a  specific  SO$(4,4)$ transformation  --- a Weyl reflection --- on the matrix of scalars for the Myers-Perry instanton.  This Weyl  reflection allows us to match the final solution rather directly to the JMaRT parameterisation upon adding charges. In section  \ref{sec:charging} we perform the charging transformations on the Weyl reflected Myers-Perry instanton matrix. The corresponding six-dimensional fields match on to the over-rotating Cveti\v{c}-Youm metric.  

We present in detail the inverse scattering construction of the Myers-Perry instanton  metric in appendix \ref{app:MPins}. 
Certain details on the construction of the SO$(4,4)$ matrix and the action of the Weyl reflection on  three-dimensional scalars are provided in appendix \ref{app:dimred}. Details on the construction of the six-dimensional fields are provided in appendix \ref{app:C2}.
A discussion on the rod structure of the Cveti\v{c}-Youm metric is presented  in appendix  
\ref{rod_structure}. The black hole and the fuzzball cases are analysed separately. 

We end with a brief discussion in section \ref{conclusions}.

%%%%%%%%%%%%%%%%%%%%%%%%%
\section{JMaRT as charged Myers-Perry instanton}
\label{JMaRT_main}

The JMaRT solutions presented in Ref.~\cite{Jejjala:2005yu} were originally obtained by starting with a large family of metrics and determining special choices of parameters that rendered the geometries smooth and horizonless. Specifically, the starting point was the general five-dimensional non-extremal solutions, derived by Cveti\v{c} and Youm~\cite{Cvetic:1996xz}, carringy two angular momenta and three independent U(1) charges, in addition to a mass parameter $M$. 
These metrics are solutions to five-dimensional supergravity theory obtained from ten-dimensional type IIB supergravity upon compactification on $T^4 \times S^1$.
While the compact $T^4$ part of the metric does not play a significant role in the JMaRT construction, the $S^1$ direction is crucial for the smoothness analysis. Therefore, the metric and matter fields are most conveniently considered as six-dimensional quantities. Our goal is to demonstrate that the JMaRT solutions can be generated in an alternative and more direct way.

%%%%%%%%%%%%%%%%%%%%%%%%%
\subsection{Myers-Perry instanton}
\label{sec:MPins}

The five-dimensional Myers-Perry instanton metric can be expressed as
\begin{flalign}
ds_{5d}^2 &= dy^2 + \frac{M}{\Sigma} \left[dy +a_1\sin^2\theta\, d\phi + a_2\cos^2\theta\, d\psi \right]^2  \nn\\
&+ (r^2-a_1^2)\sin^2\theta \, d\phi^2
+ (r^2-a_2^2)\cos^2\theta \, d\psi^2 
+ \frac{\Sigma}{\Delta}dr^2 + \Sigma\; d\theta^2,
\label{MPins:main}
\end{flalign}
where
\begin{align}
\Sigma &= r^2 -a_1^2 \cos^2 \theta - a_2^2 \sin^2\theta, &
\Delta &= r^2 \left( 1 - \frac{a_1^2}{r^2} \right) \left( 1-\frac{a_2^2}{r^2} \right) + M.
\end{align}
This is a vacuum solution of Euclidean gravity possessing three commuting Killing vector fields, namely $\partial_y, \partial_\phi$ and $\partial_\psi$, and is parametrised by the three numbers $M, a_1$ and $a_2$.
We obtain a Lorentzian metric by trivially lifting to six-dimensions through the addition of a flat time direction,
\begin{flalign}
ds_{6d}^2 =& -dt^2 + ds_{5d}^2.
\label{MPins:6d}
\end{flalign}
The line element~\eqref{MPins:main} can be easily obtained by the following analytic continuation on the Myers-Perry metric as given in Ref.~\cite{Harmark:2004rm}:
\be
\begin{array}{lll}
  a_1  & \to &    -i a_1,  \\
   a_2   & \to    & -i a_2,  \\
 t     &   \to  &    +i y,  \\
 M     &  \to    & - M.
\end{array}
\ee
%\textcolor{grey}{Metric \eqref{MPins:main}  is to be compared with the five-dimensional spatial part of Eq.~(4.13) in Ref.~\cite{Katsimpouri:2014ara}, which corresponds to the singly spinning case. Indeed, that line element is recovered by setting $a_2=0$, and redefining  $r\to\tilde{r}$ (note that $\Sigma$ becomes equal to $\tilde{f}$ in~\cite{Katsimpouri:2014ara}.)}
%\jvr{I suggest deleting this comment from this section and keeping it only in App. A.}

A standard Euclidean version of the Myers-Perry solution would not include the analytic continuation on the mass parameter, $M\to-M$~\footnote{Nevertheless, with a slight abuse of language we will continue to call metric~\eqref{MPins:main} --- and its six-dimensional uplift~\eqref{MPins:6d} --- the Myers-Perry instanton.}. While this raises questions about the regularity of such geometries, we are not concerned with the smoothness properties of this metric \textit{per se}. In section~\ref{sec:charging} below, we will add charges on top of this metric and it is the smoothness properties of the final charged metric that we will be interested in. The same approach was taken in other references, see e.g., \cite{Giusto:2007tt, Banerjee:2014hza}.

%%%%%%%%%%%%%%%%%%%%%%%%%
\subsection*{Inverse scattering construction}

The 3-parameter family of solutions~\eqref{MPins:main} can also be constructively generated from five dimensional Euclidean Schwarzschild metric by applying the Belinski-Zakharov (BZ) inverse scattering method. This procedure is detailed 
in appendix~\ref{app:MPins} and parallels the derivation of the 5D Myers-Perry metric from Schwarzschild metric in Lorenztian gravity~\cite{Pomeransky:2005sj}. One of the key points that is borne out by this construction is that the parameters must obey
\be
M < (a_1- a_2)^2.
\label{bound}
\ee
This bound arises in the JMaRT solutions as a condition ensuring that the smooth geometries are horizonless~\cite{Jejjala:2005yu}.

As is well known for the Lorentzian Myers-Perry metric, the inverse scattering procedure is not unique. The same is true for the Euclidean metric. In appendix~\ref{app:MPins} we describe one such way of generating the Euclidean solution. A brief summary is as follows. Let us recall that stationary axi-symmetric solutions of vacuum Einstein equations in five-dimensions  can  be expressed in canonical coordinates in the form~\cite{Harmark:2004rm}
\be
ds^2 = G_{ab}(\rho,z)\; dx^a dx^b+ e^{2\nu(\rho,z)}(d\rho^2+dz^2)\,, \qquad
\text{with} \quad \det G = \rho^2\,.
%\label{eq:canonicalmetric}
\ee
Note that the determinant of the Killing matrix $G_{ab}$ is positive, since we are working in Euclidean gravity.  In canonical coordinates the vacuum Einstein equations yield a decoupled set of equations for the Killing metric $G_{ab}$. These equations can be equivalently formulated as a system of first order differential equations (the Lax pair) for the so-called generating matrix. One `dresses' the generating matrix of the seed solution appropriately to obtain a new solution. 

We follow the procedure of Ref.~\cite{Pomeransky:2005sj}. We first remove a soliton and an anti-soliton with `trivial' BZ vectors from the five dimensional Euclidean Schwarzschild metric, and then add the same soliton and the anti-soliton with `nontrivial' BZ vectors. Changing the coordinates from canonical to  more standard radial coordinates, and choosing convenient names for the parameters added through the BZ vectors,  we obtain the metric \eqref{MPins:main} together with the bound \eqref{bound}. A step-by-step description of the procedure is presented in appendix~\ref{app:MPins}.

%%%%%%%%%%%%%%%%%%%%%%%%%
\subsection*{Shifted coordinates}

For the ensuing discussion the following coordinates are more useful to work with. These coordinates allow to match rather directly the charged version of the Myers-Perry instanton to the over-rotating Cveti\v{c}-Youm metric. The coordinate transformation is 
\bea
r^2 &\longrightarrow& r^2 + a_1^2 + a_2^2 - M,  \label{shift1}\\
\theta &\longrightarrow& \frac{\pi}{2} -\theta. \label{shift2}
\eea
Along with these coordinate shifts, we also interchange  coordinates $\phi$ and $\psi$ and  names of the rotation parameters $a_1$ and $a_2$:
\bea
\phi &\longleftrightarrow& \psi, \\
a_1 &\longleftrightarrow& a_2. 
\eea
The resulting metric reads
\begin{flalign}
ds_{6d}^2 =& -dt^2 + dy^2 + \frac{M}{\tilde\Sigma} \left[dy +a_1\sin^2\theta\, d\phi + a_2\cos^2\theta\, d\psi \right]^2  \nn\\
&+ (r^2+a_2^2-M)\sin^2\theta \, d\phi^2
+ (r^2+a_1^2-M)\cos^2\theta \, d\psi^2 
+ \frac{\tilde\Sigma}{\tilde\Delta}dr^2 + \tilde\Sigma\; d\theta^2,
\label{MPinsShifted}
\end{flalign}
where
\bea
\tilde \Sigma  &=& r^2 + a_1^2 \sin^2 \theta + a_2^2 \cos^2\theta - M, \label{tSig}\\
\tilde \Delta &=&
%& =: & \frac{g(r)}{r^2} =  
r^2 \left( 1 + \frac{a_1^2}{r^2} \right)\left( 1 + \frac{a_2^2}{r^2}\right) - M \label{tDel}.
\eea

\subsection{Dimensional reduction to 3d and Weyl reflection}
\label{sec:Weyl}

As our next step we will apply a solution generating technique based on three-dimensional duality symmetries on the Myers-Perry instanton metric \eqref{MPinsShifted}. Thus, we begin by dimensionally reducing down to three dimensions.

The six-dimensional truncation of IIB theory on $T^4$ that we work with is
\beq
\mL_6=R_6 \star_6 1-\half \star_6 d\Phi\wedge d\Phi-\half e^{\sq{2}\Phi} \star_6 F_{[3]}\wedge F_{[3]} ,
\label{Lagrangian_main_text}
\eeq 
where the field strength $F_{[3]}=dC_{[2]}$ comes from the Ramond-Ramond sector of the ten-dimensional IIB theory. The six-dimensional metric \eqref{MPinsShifted} is viewed as a solution of theory \eqref{Lagrangian_main_text}, specifically a solution with trivial dilaton $\Phi$ and two-form field $C_{[2]}$.

%%%%%%%%%%%%%%%%%%%%%%%%%
\subsection*{Three-dimensional dualities}

Upon dimensional reduction a large number of gravity and supergravity theories become gravity coupled to form-fields and non-linear sigma models. Such non-linear sigma models are maps from a lower-dimensional base space to a target space. The target space is generally a coset $\mathrm{G}/\mathrm{K}$. The group G is the group of global isometries of the target space. The group K is the isotropy subgroup of the target space -- a subgroup of G. The symmetry group  G of a sigma model can be used to generate new solutions of the higher-dimensional gravity theory by applying a group transformation to a coset representative of a seed solution. 

These techniques become particularly powerful when the reduction is performed down to three dimensions. In three dimensions all higher dimensional form fields can be dualized to scalars. As a result the symmetry groups become significantly enhanced, and one has at ones disposal a rich solution generating technique. Further richness comes from changing the details of the dimensional reduction. For example, by changing the order of the timelike reduction within the whole sequence of reductions, one can change the denominator subgroup.  

These techniques have been presented at several places in the literature, see e.g., \cite{Youm:1997hw}; we will not review it here. We refer the reader to appendix~\ref{app:dimred} for some more details and notation. The key quantity in this method to work with is a matrix $\cM$ that encodes all three-dimensional scalars.
These are obtained by performing a sequence of Kaluza-Klein reductions down to 3d, together with the dualisation of the one-forms that are left over.
The matrix $\cM$ belongs to the coset $\mathrm{G}/\mathrm{K}$.

For the theory  \eqref{Lagrangian_main_text} the coset model is 
\be
\frac{\mathrm{SO}(4,4)}{\mathrm{SO}(2,2) \times \mathrm{SO}(2,2)},
\ee
where the embedding of the denominator subgroup  in the numerator group depends on the details of the dimensional reduction.
The specific ordering of the Kaluza-Klein reductions we adopted was over $y$, $\phi$, and $t$, respectively.
Group transformations with elements belonging to the denominator subgroup act as
\be
\cM \to g^{-1} \ \cM \ g, \qquad \qquad \mbox{for} \qquad g \ \ \in  \ \ \mathrm{SO}(2,2) \times \mathrm{SO}(2,2).
\ee
Thus, from the metric \eqref{MPinsShifted} we construct the SO(4,4) matrix $\cM$, roughly by exponentiating the various group generators --- each generator being weighted by one of the 3d scalars --- and multiplying them all together. The group SO(4,4) has dimension 28. The Cartan subalgebra is spanned by four generators, denoted $H_\Lambda$, with $\Lambda=0,\dots,3$. The remaining 24 generators are broken into `positive' ($E_\Lambda, E_{q_\Lambda}, E_{p^\Lambda}$) and `negative' ($F_\Lambda, F_{q_\Lambda}, F_{p^\Lambda}$) elements and the number of available 3d scalars (sixteen) matches the number of Cartan plus positive generators. More details are given in appendix~\ref{app:dimred}. We adopted the same basis for the $\so(4,4)$ algebra as the one defined in Refs.~\cite{Bossard:2009we, Virmani:2012kw}.

%%%%%%%%%%%%%%%%%%%%%%%%%
\subsection*{Weyl reflection}

On the resulting matrix $\cM$ we act with the following group element 
\be
g_w = \exp\left[ i \frac{\pi}{2} K_{q_2} \right] \exp\left[ i \frac{\pi}{2} K_{q_3} \right], 
\label{Weyl}
\ee
as
\be
\cM_w = g_w^{-1} \cM g_w.
\ee
Here, we have defined $K_{q_\Lambda}:=E_{q_\Lambda}-{E_{q_\Lambda}}^\sharp$, where the symbol ${}^\sharp$ denotes the generalised transpose [see appendix~\ref{app:dimred} below Eq.~\eqref{MVV}].
Although complex numbers appear in  definition \eqref{Weyl}, it can be checked by direct inspection that the resulting matrix is real and indeed belongs to the denominator $\mathrm{SO}(2,2) \times \mathrm{SO}(2,2)$ subgroup of the numerator $\mathrm{SO}(4,4)$ group. We follow the $\so(4,4)$ Lie algebra conventions of \cite{Bossard:2009we, Virmani:2012kw}.

In the numerator $\mathrm{SO}(4,4)$, $g_w$ is a Weyl reflection. Of particular interest is the action of this transformation on the Euclidean gravity truncation to which the metric \eqref{MPinsShifted}  belongs. As is discussed in  detail in appendix  \ref{app:dimred}, its action changes the truncation from Euclidean five-dimensional vacuum gravity to Lorentzian five-dimensional vacuum gravity. The bound \eqref{bound} on the parameters  does not change. The resulting matrix $\cM_w$ can be thought of as describing `over-rotating' Lorentzian Myers-Perry metric. This needs to be contrasted with the inverse scattering construction of  the Lorentzian Myers-Perry metric, e.g., as presented in \cite{Pomeransky:2005sj}, where the bound \eqref{bound}  cannot be fulfilled with real pole positions in the dressing transformations. 
A very similar transformation was used in \cite{Katsimpouri:2014ara}. However, details are not identical. 

Of course, one could have taken directly, as a starting point, the `over-rotating' Myers-Perry solution and then charge it up as we will do next. But by following this longer route we emphasise that the JMaRT smooth solutions can be systematically constructed from gravitational instantons.

%%%%%%%%%%%%%%%%%%%%%%%%%%
\subsection{Charging transformations and 6d fields}
\label{sec:charging}
On the resulting matrix $\cM_w$ we act with a charging transformation that adds three electric charges. We choose names for the charging parameters so that the final answer conforms to the JMaRT notation. The charging transformation is
\be
g_\rom{c} = \exp\left[ \delta_p K_{q_1} \right] \exp\left[ - \delta_1 K_{q_2} \right] \exp\left[  \delta_5 K_{q_3} \right], 
\ee
acting as
\be
\qquad \cM_\rom{final} = g_c^{-1} \cM_w g_c.
\label{charging}
\ee

We read scalars from the matrix $\cM_\rom{final}$ and build the metric, dilaton, and the C-field in six-dimensions. We find an answer identical to the fields given in reference \cite{Jejjala:2005yu}. Certain details on the construction of the six-dimensional fields are provided in appendix \ref{app:C2}.  For completeness, and for use in appendices, we write the final fields here. The six-dimensional Einstein frame metric reads
\begin{eqnarray} \label{3charge}
ds^2_{6d}&=&-\frac{f}{\sqrt{\tilde{H}_{1} \tilde{H}_{5}}}(
dt^2 - dy^2) +\frac{M}{\sqrt{\tilde{H}_{1}
\tilde{H}_{5}}} (s_p dy - c_p
dt)^2 \nonumber \\ &&+\sqrt{\tilde{H}_{1} \tilde{H}_{5}}
\left(\frac{ r^2 dr^2}{ (r^2+a_{1}^2)(r^2+a_2^2) - Mr^2}
+d\theta^2 \right)\nonumber \\ &&+\left( \sqrt{\tilde{H}_{1}
\tilde{H}_{5}} - (a_2^2-a_1^2) \frac{( \tilde{H}_{1} + \tilde{H}_{5}
-f) \cos^2\theta}{\sqrt{\tilde{H}_{1} \tilde{H}_{5}}} \right) \cos^2
\theta d \psi^2 \nonumber \\ && +\left( \sqrt{\tilde{H}_{1}
\tilde{H}_{5}} + (a_2^2-a_1^2) \frac{(\tilde{H}_{1} + \tilde{H}_{5}
-f) \sin^2\theta}{\sqrt{\tilde{H}_{1} \tilde{H}_{5}}}\right) \sin^2
\theta d \phi^2 \nonumber \\ && +
\frac{M}{\sqrt{\tilde{H}_{1} \tilde{H}_{5}}}(a_1 \cos^2 \theta
d \psi + a_2 \sin^2 \theta d \phi)^2 \nonumber \\ &&
+ \frac{2M \cos^2 \theta}{\sqrt{\tilde{H}_{1} \tilde{H}_{5}}}[(a_1
c_1 c_5 c_p -a_2 s_1 s_5 s_p) dt + (a_2 s_1
s_5 c_p - a_1 c_1 c_5 s_p) dy ] d\psi \nonumber \\ 
&&+\frac{2M \sin^2 \theta}{\sqrt{\tilde{H}_{1} \tilde{H}_{5}}}[(a_2
c_1 c_5 c_p - a_1 s_1
s_5 s_p) dt + (a_1
s_1 s_5 c_p - a_2 c_1 c_5 s_p) dy] d\phi,
\end{eqnarray}
where
\begin{eqnarray} 
\tilde{H}_{i}=f+M\sinh^2\delta_i, \quad
f=r^2+a_1^2\sin^2\theta+a_2^2\cos^2\theta,
\end{eqnarray}
and $c_i = \cosh \delta_i$, $s_i = \sinh \delta_i$.  The six-dimensional two-form is given by

\begin{eqnarray} \label{RR}
C_2 &=& - \frac{M s_1 c_1}{\tilde H_1} dt \wedge dy -
  \frac{M s_5 c_5}{\tilde H_1} (r^2 + a_2^2 + M
  s_1^2) \cos^2 \theta d\psi \wedge d\phi \\
&&   +\frac{M \cos^2 \theta}{\tilde H_1} \left[ (a_2 c_1 s_5 c_p - a_1 s_1 c_5  s_p) dt + (a_1 s_1 c_5 c_p - a_2 c_1 s_5 s_p) dy \right] \wedge d\psi \nn \\ && + \frac{M \sin^2 \theta}{\tilde H_1} \left[  (a_1 c_1 s_5 c_p - a_2 s_1 c_5 s_p) dt  + (a_2 s_1 c_5 c_p - a_1 c_1 s_5 s_p) dy \right] \wedge d \phi, \nn 
\end{eqnarray}
and finally the six-dimensional dilaton $\Phi$, cf.~\eqref{Lagrangian}, reads
\begin{equation}
e^{2\sqrt{2}\Phi} = \frac{\tilde H_1}{\tilde H_5}.
\end{equation}
A discussion of the rod structure for this metric is presented in appendix~\ref{rod_structure}.

%%%%%%%%%%%%%%%%%%%%%%%%
%%%%%%%%%%%%%%%%%%%%%%%%
\section{Conclusions}
\label{conclusions}

In this paper we have presented an alternative and more direct (inverse-scattering based) construction 
of the over-rotating Cveti\v{c}-Youm metric. We have generalized --- and at the same time simplified --- the construction of \cite{Katsimpouri:2014ara}. Certain further restrictions on the parameters of the resulting 6d fields  give rise to a discrete family of non-extremal smooth bound states  of the D1-D5-P system \cite{Jejjala:2005yu}.

Another objective of this work was to emphasise the idea that the over-rotating Cveti\v{c}-Youm metric can be viewed as a charged version of the Myers-Perry instanton metric. 
Indeed, this picture is strongly suggested by the similarities between the rod structures of the two metrics. Although the Cveti\v{c}-Youm geometry is not a vacuum solution, from the metric alone one can still define a rod structure and this was presented in appendix~\ref{rod_structure}.

More generally, one may hope that adding appropriate charges to gravitational instantons might lead to a  class of non-supersymmetric fuzzballs. 
It will be very exciting if this circle of ideas can be pushed further to construct a  class of multi-bubble non-extremal fuzzball solutions.  Given the remarkable success that the inverse scattering method has had with black rings, we expect that progress should be possible on ``three-center''  non-extremal solutions.
This may be achieved by generalising the present study by taking a (yet unknown) Euclidean black ring as the starting point for the charging transformation.
It will also be interesting to understand the recent construction of \cite{Bena:2015drs}  from the point of view pursued in this paper.

%%%%%%%%%%%%%%%%%%%%%%%%
%%%%%%%%%%%%%%%%%%%%%%%%
\subsection*{Acknowledgements}
We have benefitted from our discussions with Iosif Bena, Geoffrey Comp\`ere, and Axel Kleinschmidt. We also thank David Turton for collaboration at very early stages of this work.
JVR thanks the STAG Research Center at the University of Southampton for hospitality during the completion of this work.
J.V.R. acknowledges financial support from the European Union's Horizon 2020 research and innovation programme under the Marie Sklodowska-Curie grant agreement No REGMat-656882.
Funding for this work was partially provided by the Spanish MINECO under project FPA2013-46570-C2-2-P.

%\newpage

%%%%%%%%%%%%%%%%%%%%%%%%
%%%%%%%%%%%%%%%%%%%%%%%%
\appendix
\section{Inverse scattering construction of the Myers-Perry instanton}
\label{app:MPins} 

In the interest of providing a complete derivation of the JMaRT solutions, we present in this Appendix all the details necessary to generate the Myers-Perry instanton from the Euclidean Schwarzschild solution using the Inverse Scattering Method (ISM). As is well known, the procedure is not uniquely determined. Below we describe, step by step, one such way of generating this solution.
To set the context, and also to fix some notation, we begin by offering a very concise account of the formalism.

\subsection*{Overview of the procedure}

Recall that solutions of the vacuum Einstein equations in $D=5$ dimensions, $R_{\mu\nu}=0$, that are both stationary and (doubly-)axially symmetric (thus possessing $D-3$ commuting Killing vector fields) can always be expressed in canonical coordinates in the form~\cite{Harmark:2004rm}\footnote{Since we are working in the Euclidean section, the determinant of the {\it Killing matrix} $G_{ab}$ is positive. For Lorentzian solutions we would have an extra minus sign on the far right hand side of~\eqref{eq:canonicalmetric}.}
\be
ds^2 = G_{ab}(\rho,z)\; dx^a dx^b+ e^{2\nu(\rho,z)}(d\rho^2+dz^2)\,, \qquad
\text{with} \quad \det G = \rho^2\,.
\label{eq:canonicalmetric}
\ee
In these coordinates the vacuum Einstein equations yield a decoupled elliptic PDE for the Killing metric $G_{ab}$. This can be equivalently formulated as a system of {\em first order linear} equations (the Lax pair) for the so-called generating matrix, which depends on an additional variable (the spectral parameter). A linear transformation on this generating matrix --- in standard terminology, one refers to it getting {\it dressed} --- takes us to a new solution of the same field equations. Under the assumption of a linear transformation that adds only simple poles in the spectral parameter complex plane (i.e. a {\it solitonic} transformation) the whole procedure reduces to a sequence of algebraic calculations~\cite{BZ}. The determination of the {\em conformal factor} $e^{2\nu}$ can be straightforwardly accomplished by a line integral once the Killing matrix is found. Nevertheless, even this can be sidestepped since the conformal factor of the new solution can be directly obtained from that of the seed solution via another simple algebraic evaluation.

%%%%%%%%%%%%%%%%%%%%%%%%%%%
\subsection*{Details of the ISM construction}

After this lightening review of the ISM, we now move on to the construction of the 5D Euclideanized Myers-Perry geometry, closely following Pomeransky's derivation of 5D Lorentzian Myers-Perry \cite{Pomeransky:2005sj}. %\jvr{Refer to the file ``MPInstanton\_2soliton.nb'' for explicit computations.} 
This instanton can be connected with the zero-charge JMaRT solution by later adding a flat timelike direction~\cite{Katsimpouri:2014ara}. The construction proceeds as follows:
\begin{enumerate}
  \item The starting point is the diagonal metric corresponding to 5D Euclidean Schwarzschild, which is written in the form~\eqref{eq:canonicalmetric}, with $G=G_0$ and $\nu=\nu_0$ (the ``$0$'' in the subscript refers to the seed solution),
\be
(G_0)_{ab}=\text{diag}\left\{\frac{\mu_1}{\mu_2},\mu_2,\frac{\rho^2}{\mu_1}\right\}\,.
\label{eq:seedKM}
\ee
The rod diagram for such a solution is displayed in Fig.~\ref{roddiagram} ($\Omega_\phi$ and $\Omega_\psi$ must be set to zero). The Killing sector is parametrized by coordinates $(y,\phi,\psi)$ and the solitons and anti-solitons are defined, respectively, by
\be
\mu_i = \sqrt{\rho^2 + (z-b_i)^2} - (z-b_i)\,, \qquad
\overline{\mu}_i = -\sqrt{\rho^2 + (z-b_i)^2} - (z-b_i)\,.
\ee
They satisfy $\mu_i\overline{\mu}_i=-\rho^2$.

\begin{figure}[t!]
\centering{
\includegraphics[width=9cm]{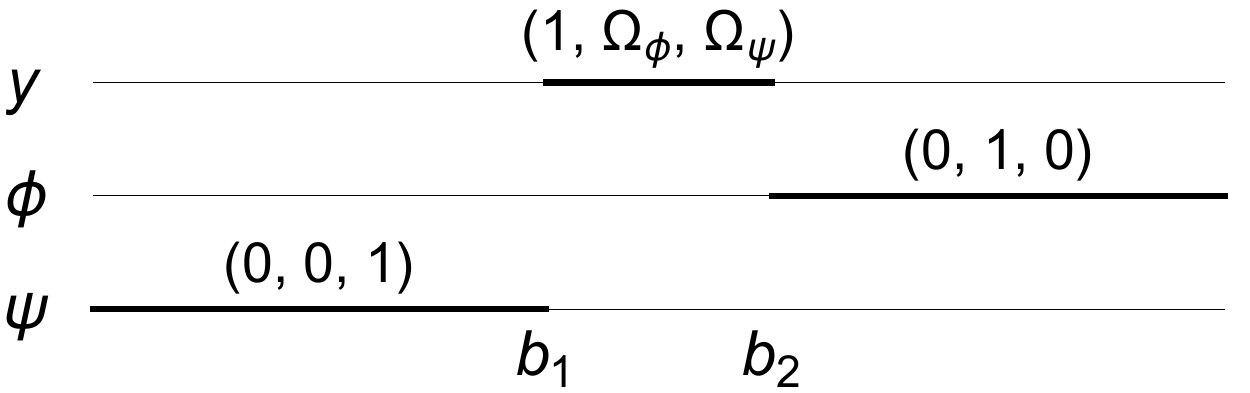}}
\bigskip
\caption{Rod diagram for the 5D Euclidean Myers-Perry geometry. The direction for each rod is indicated above the corresponding segment. The rod diagram for the seed solution (Euclidean Schwarzschild) is trivially obtained by setting both ``angular velocities'' $\Omega_\phi$ and $\Omega_\psi$ to zero. The points $b_1$ and $b_2$ indicate turning points where regularity of the solution has to be checked explicitly.}
\label{roddiagram}
\end{figure}

  \item The conformal factor for this seed is algorithmically determined by following the procedure described in Ref.~\cite{Izumi:2007qx},
\be
e^{2\nu_0}=k^2\frac{\mu_2 \left(\mu_1 \mu_2+\rho ^2\right)}{\left(\mu_1^2+\rho ^2\right) \left(\mu_2^2+\rho ^2\right)}\,.
\ee
The multiplicative constant $k$ can be fixed by requiring asymptotic flatness. 
  \item From the seed Killing matrix~\eqref{eq:seedKM} we\footnote{This step is necessary in $D>4$ to ensure that the final solution satisfies the constraint $\det G=\rho^2$ in Eq.~\eqref{eq:canonicalmetric}. Refer to e.g. Refs.~\cite{Emparan:2008eg, Elvang:2007rd, Rocha:2013qya} for concise accounts of the details of the ISM procedure.}:
  \begin{enumerate}
    \item remove a soliton at $z=b_1$ with trivial BZ vector $m_0^{(1)}=(0,0,1)$, which amounts  to dividing $G_{\psi\psi}$ by $-\rho^2/\mu_1^2$;
    \item remove an anti-soliton at $z=b_2$ with trivial BZ vector $m_0^{(2)}=(0,1,0)$, which amounts  to dividing $G_{\phi\phi}$ by $-\mu_2^2/\rho^2$;
    \item multiply the whole matrix by a factor $-\mu_2/\mu_1$, for convenience.
  \end{enumerate}
  The Killing matrix thus obtained is
\be
(G_0')_{ab}=\text{diag}\left\{-1,\overline{\mu_1},\mu_2\right\}\,.
\label{eq:interKM}
\ee
  This will serve as the seed for the next solitonic transformation.
  \item Now we add the (anti-)solitons that we removed previously but with nontrivial BZ vectors. Namely, we:
  \begin{enumerate}
    \item add a soliton at $z=b_1$ with BZ vector $m_0'^{(1)}=(A_1,0,C_1)$;
    \item add an anti-soliton at $z=b_2$ with BZ vector $m_0'^{(2)}=(A_2,B_2,0)$.
  \end{enumerate}
  At this stage we have obtained a new Killing matrix. 
  Clearly, if we set $A_1=A_2=0$ and $C_1=B_2=1$ (and rescale to revert step 3.(c)) this just undoes the previous step and so we must retrieve the original solution. It is the presence of non vanishing coefficients $A_i$ that mixes $y$ (Euclidean time) and angular components. In the Lorentzian picture this would correspond to turning on angular velocities.
  \item Rescale again the Killing matrix (multiply it by $-\mu_1/\mu_2$) to undo the scaling of step 3.(c). This yields a physical metric satisfying the constraint $\det G=\rho^2$. However, the orientation of the rods is non standard: the solitonic transformation performed to mix $y$ direction and angular components simultaneously rotated the directions of the outermost rods. So an analysis of the rods' orientation must be done at this point, which we turn to next.
  \item It is convenient to set $b_1=-b_2=-\alpha$, with $\alpha > 0$, without loss of generality\footnote{The metric~\eqref{eq:canonicalmetric} with $G$ and $e^{2\nu}$ depending on $z$ only through the combinations $\mu_i$ is invariant under simultaneous shifts of the $z$ coordinate and the $b_i$ parameters.}. A rod structure analysis reveals that:
  \begin{enumerate}
    \item the rightmost rod (rod 3: $\rho=0, \; z>\alpha$) has orientation $\left(-\frac{4\alpha A_2}{B_2}, 1, \frac{4\alpha A_1 A_2}{B_2 C_1}\right)$;
    \item the leftmost rod (rod 1: $\rho=0, \; z<-\alpha$) has orientation $\left(-\frac{4\alpha A_1}{C_1}, \frac{4\alpha A_1 A_2}{B_2 C_1}, 1\right)$.
  \end{enumerate}
  As a useful check, we confirm that a trivial solitonic transformation ($A_i=0$) does not change the direction of the rods.
  \item The linear transformation $G \to \Lambda^{T} G \Lambda$, with
\be
\Lambda=\left(
\begin{array}{ccc}
 1 & -4 A_2 C_1 \alpha  & -4 A_1 B_2 \alpha  \\
 0 & B_2 C_1 & 4 A_1 A_2 \alpha  \\
 0 & 4 A_1 A_2 \alpha  & B_2 C_1 \\
\end{array}
\right)\,,
\ee
brings us back to standard orientation (so that rod 1 and rod 3 are aligned with directions $(0,0,1)$ and $(0,1,0)$, respectively). In the process the finite middle rod 2 acquires direction $(1,\Omega_\phi,\Omega_\psi)$, where
\be
\Omega_\phi = \frac{A_2}{C_1(4 \alpha A_2^2 - B_2^2)}\,, \qquad
\Omega_\psi = \frac{A_1}{B_2(4\alpha A_1^2-C_1^2)}\,.
\ee
  We have thus generated the Euclidean Myers-Perry solution.
\end{enumerate}

%%%%%%%%%%%%%%%%%%%%%%%%%%%
\subsection*{Final metric in convenient coordinates}

The solution as obtained above (but not explicitly shown), written in canonical coordinates $(\rho,z)$, is not particularly illuminating and it is desirable to express it in a more compact form.
One useful system is the choice of prolate spherical coordinates $(u,v)$, related with the canonical coordinates through
\be
\rho=\alpha \sqrt{\left(u^2-1\right) \left(1-v^2\right)}\,, \qquad
z=\alpha  u v\,,
\label{canonicalpro}
\ee
where $u\geq1$ and $-1\leq v\leq1$.

Besides changing coordinates, it is also convenient to redefine the parameters. The parameters characterising the solution are $\alpha, A_1/B_2, A_2/C_1$. The dependence of the solution only on the ratios $A_1/B_2$ and $A_2/C_1$ is a consequence of the invariance of the ISM procedure under rescalings of the BZ vectors, $m_0^{(i)} \to \lambda_i m_0^{(i)}$, with $\lambda_i\neq0$.
Following Pomeransky~\cite{Pomeransky:2005sj} we fix the normalisation
\be
B_2^2 C_1^2-16 \alpha^2 A_1^2 A_2^2=1\,,
\ee
which simplifies intermediate steps of the calculation. Then we define
\bea
M &=& -4\alpha \left(4\alpha A_1^2-C_1^2\right) \left(4\alpha A_2^2-B_2^2\right)\,,\\
a_1 &=& 4 \alpha  A_2 C_1\,,\\
a_2 &=& 4 \alpha  A_1 B_2\,.
\eea
%\jvr{There was a typo in Eq.(A.10): the signs in front of $C_1^2$ and $B_2^2$ were wrong. In the notebook they are flipped. Actually, as is was before it would imply $M<0$.}
Note that $\alpha, a_1, a_2$ and $M$ are not all independent since they satisfy
\be
M = a_1^2+a_2^2 - 2 \sqrt{4\alpha^2+a_1^2 a_2^2}\,.
\label{Ma1a2alpha}
\ee
The requirement that $\alpha$ should be real and positive, i.e., the location of rod endpoints are as described above, implies
\be
M < (a_1- a_2)^2.
\ee

After applying all these transformations we obtain the Euclidean Myers-Perry solution in prolate spherical coordinates.
We present the final metric in a different set of coordinates, $(r,\theta)$, closely related to the coordinates used in the Cveti\v{c}-Youm and JMaRT papers. They are related with $(u,v)$ through
\be
\alpha^2 \left(u^2-1\right) \left(1-v^2\right)=\frac{r^2}{4} \Delta \sin^2(2\theta)\,, \qquad
\alpha u v=\frac{r^2}{2} \left(1-\frac{a_1^2+a_2^2-M}{2 r^2}\right) \cos(2\theta)\,,
\label{prospherical}
\ee
where
\be
\Delta \equiv r^2 \left(1-\frac{a_1^2}{r^2}\right) \left(1-\frac{a_2^2}{r^2}\right)+M\,.
\ee

It is convenient to introduce the following combination:
\be
\Sigma = r^2 - a_1^2 \cos^2\theta - a_2^2 \sin^2\theta.
\ee
In terms of these new coordinates the metric is expressed as
\begin{flalign}
ds^2 =& dy^2 + \frac{M}{\Sigma} \left[dy +a_1\sin^2\theta\, d\phi + a_2\cos^2\theta\, d\psi \right]^2  \nn\\
&+ (r^2-a_1^2)\sin^2\theta \, d\phi^2
+ (r^2-a_2^2)\cos^2\theta \, d\psi^2 
+ \frac{\Sigma}{\Delta}dr^2 + \Sigma\; d\theta^2.
\label{MPins}
\end{flalign}
This metric is to be compared with the five-dimensional spatial part of Eq.~(4.13) in Ref.~\cite{Katsimpouri:2014ara}, which corresponds to the singly spinning case. Indeed, that line element is recovered by setting $a_2=0$, and redefining  $r\to\tilde{r}$ (note that $\Sigma$ becomes equal to $\tilde{f}$ in~\cite{Katsimpouri:2014ara}.)

%%%%%%%%%%%%%%%%%%%%%%%%%
\section{From 6d to 3d and back}
\label{app:dimred}
In this appendix we present some details on 6d to 3d reduction. We follow conventions of \cite{Virmani:2012kw}. We focus on details complementary to what is already presented in that reference.
\subsection*{Notation}
A well known truncation of IIB supergravity on T$^4$ has  6D Lagrangian 
\beq
\mL_6=R_6 \star_6 1-\half \star_6 d\Phi\wedge d\Phi-\half e^{\sq{2}\Phi} \star_6 F_{[3]}\wedge F_{[3]} ,
\label{Lagrangian}
\eeq 
where the field strength $F_{[3]}=dC_{[2]}$ comes from the RR sector of the ten-dimensional IIB theory. As discussed in appendix A of \cite{Virmani:2012kw} upon dimensional reduction on a spacelike circle the 6D theory reduces to the U(1)$^3$ supergravity in 5D. The reduction ansatz for the metric and the 3-form field strength are 
\bea
ds_6^2&=& e^{-\sq{\f{3}{2}}\Psi}(dz_6+A^1_{[1]})^2+e^{\f{\Psi}{\sq{6}}}ds_5^2 ,\\
F_{[3]}&=& F_{[3]}^{5d}+dA^2_{[1]}\wedge (dz_6+A^1_{[1]}) ,
\eea
with \beq
F_{[3]}^\rom{(5d)}=dC_{[2]}^\rom{(5d)}-dA^2_{[1]}\wedge A^1_{[1]}  .
\eeq
After dualizing $C_{[2]}^\rom{(5d)}$ to a vector $A^3_{[1]}$ in 5D using the method of Lagrange multipliers, the triality structure of U(1)$^3$ supergravity becomes manifest.

Now we have obtained two scalars in five-dimensions, namely $\Psi$ and $\Phi$. 
We parameterise the U(1)$^3$ supergravity scalars as 
\be
h^1 =e^{\sq{\f{2}{3}}\Psi},\quad h^2=e^{-\sq{\f{1}{6}}\Psi-\sq{\f{1}{2}}\Phi},\quad h^3=e^{-\sq{\f{1}{6}}\Psi+\sq{\f{1}{2}}\Phi} ,
\label{dilatonic_scalars}
\ee
which manifestly satisfy $h^1 h^2 h^3=1$.
Further dimensional reduction along a spacelike direction with the ansatz
\bea
ds_5^2=f^2 (d z_5+\check{A}^0_{[1]})^2+f^{-1}ds_4^2 ,\\
A^I_{[1]}=\check{A}^I_{[1]}+\ch^I (dz_5+\check{A}^0_{[1]}),
\eea
gives rise to the $N=2$ STU model in 4D. 
The scalars $\chi^I$ and $h^I$ combine to form complex scalars of the STU theory $z^I=-\chi^I+if h^I\equiv x^I+iy^I$.

Further dimensional reduction over a timelike direction gives an $\mathrm{SO}(4,4)/(\mathrm{SO}(2,2) \times \mathrm{SO}(2,2))$ coset model. The ansatz for this reduction step is
\bea
ds_4^2 &=&-e^{2U}(dt+\om_3)^2 +e^{-2U} ds_3^2 , \\
\check{A}^\La_{[1]} &=& A_3^{\La}+\ze^{\La}(dt+\om_3),
\eea
where $\om_3$ and $A_3^{\La}$ are 1-forms in 3D and $\La=0,\dots,3$. We dualise these vectors in 3D to scalars using a similar Lagrange multiplier method as mentioned before. The duality relations are
\beq
-d\ti{\ze}_{\La}=e^{2U}(\mathrm{Im} \, N)_{\La\Si} \star_3(dA_3^{\Si}+\ze^{\Si}d\om_3)+(\mathrm{Re} \, N)_{\La\Si}d\ze^{\Si} ,
\eeq
and
\beq
-d\si=2e^{4U} \star_3 d\om_3-\ze^{\La}d\ti{\ze}_{\La}+\ti{\ze}_{\La}d\ze^{\La} , 
\eeq
where $\ti{\ze}_{\La}$ and $\si$ are pseudo-scalars dual to $A_3^{\La}$ and $\om_3$ respectively. 
The $\mathrm{Re} \, N$ and $\mathrm{Im} \, N$ are the real and imaginary parts of the period matrix $N$ of the STU theory and they are constructed out of the $\chi^I$'s and $h^I$'s, respectively.

Therefore,  in 3D we have a total of sixteen scalars 
\beq
\varphi^a=\{U,z^I, \bar{z}^I,\ze^{\La},\ti{\ze}_{\La},\si\},
\eeq  
parameterising an  $\mathrm{SO}(4,4)/(\mathrm{SO}(2,2) \times \mathrm{SO}(2,2))$ coset  model.  
Further details on this set-up can be found in appendix A of \cite{Virmani:2012kw}, where conventions for the $\so(4,4)$ Lie algebra are also given. The resulting 3D Lagrangian is 
\beq
\mL_3=R_3 \star_3 1-\half G_{ab} \star_3 d\varphi^a \wedge  d\varphi^b .
\eeq
The whole point of the cumbersome procedure described above was to reduce the theory to such a sigma model.

If we perform the first dimensional reduction over a timelike direction and the following reductions over spacelike directions we  get a different  $\mathrm{SO}(4,4)/(\mathrm{SO}(2,2) \times \mathrm{SO}(2,2))$ coset model. One can take other combinations as well. Such reductions are used in different contexts, see \cite{Sahay:2013xda, Katsimpouri:2014ara}.

The scalar coset space can be parameterised in the Iwasawa gauge by the coset element
\beq
\mV=e^{-UH_0}\cdot \left( \prod\limits_{I=1,2,3} e^{-\half (\log{ y^I})H_I}\cdot e^{-x^I E_I} \right) \cdot e^{-\ze^{\La}E_{q_{\La}}-\ti{\ze}_{\La}E_{p^{\La}} }\cdot e^{-\half \si E_0}.
\eeq
The matrix $\cM$ is defined as 
\beq
\cM=\mV^{\sharp}\mV ,
\label{MVV}
\eeq
where $\theta^{\sharp}= \eta^{\prime} \theta^T \eta^{\prime^{-1}} $ for all $\theta\in \so (4,4)$ and $\eta^{\prime}=$diag$(-1,-1,1,1,-1,-1,1,1)$ is invariant under the action of the maximal subgroup $\mathrm{SO}(2,2) \times \mathrm{SO}(2,2)$.

\subsection*{Scalars and some relations from matrix $\cM$}
We define a matrix $\cN$ that conveniently encodes all one-forms in three dimensions, $\cN = \cM^{-1} d \cM$.  Under group transformation the matrix $\cN$  transforms as $\cN \to g^{-1} \cN g$. From this matrix one can extract duals of one forms \cite{Chow:2014cca} as follows,
\bea
\star_3 d \omega_3 &=& \cN_{74}, \\
\star_3 d A^0_3 &=& \cN_{71}, \\
\star_3 d A^1_3 &=& \cN_{81}, \\
\star_3 d A^2_3 &=& \cN_{76}, \\
\star_3 d A^3_3 &=& \cN_{72}. 
\eea
Having obtained $\star_3 d \omega_3$ one can straightforwardly integrate to construct $\omega_3$. This procedure is emphasised in references \cite{Gal'tsov:2008sh, Chow:2014cca}\footnote{We thank Geoffrey Comp\`ere for discussions on this point and for sharing some of his notes with us.} for STU supergravity. For minimal supergravity it was noted in \cite{Compere:2009zh}, though in that set-up it did not bring much technical advantage. For STU theory this procedure indeed simplifies calculations.

The remaining three-dimensional scalars are determined directly from the matrix $\cM$.  There are many ways to extract scalars from the matrix $\cM$. Among others, we have found the following equations useful  \cite{Chow:2014cca}: 
\bea
\expe{4U} & = & \frac{1}{\cM_{3 3} \cM_{4 4} - \cM_{3 4}^2}, \label{Ueqn}
\eea
\bea
\zeta^0 & =& \expe{ 4U} \left( \cM_{3 1} \cM_{3 4} - \cM_{4 1} \cM_{3 3} \right), \\
%%%
\zeta^1 & =& \expe{4U} \left(\cM_{3 1} \cM_{4 4} - \cM_{4 1} \cM_{3 4} \right), \\
%%%
\zeta^2 & =& \expe{4U} \left( \cM_{6 4} \cM_{3 3} - \cM_{6 3} \cM_{3 4} \right), \\
%%%
\zeta^3 & = & \expe{4U} \left(\cM_{3 2} \cM_{3 4} - \cM_{4 2} \cM_{3 3}\right), 
\eea
%%%
\bea
x_1  &=& \frac{\cM_{3 4}}{\cM_{3 3}}, \\
\fr{x_2}{y_2 y_3} & = & \cM_{1 6} + \expe{4 U} (\cM_{3 4} \cM_{4 1} \cM_{6 3} + \cM_{3 1} \cM_{3 4} \cM_{6 4} - \cM_{3 1} \cM_{4 4} \cM_{6 3} - \cM_{3 3} \cM_{4 1} \cM_{6 4}),  \nn \\
%%%
\fr{x_3}{y_2 y_3} & =& \cM_{1 2} + \expe{4 U} (\cM_{3 1} \cM_{3 2} \cM_{4 4} + \cM_{3 3} \cM_{4 1} \cM_{4 2} - \cM_{3 1} \cM_{3 4} \cM_{4 2} - \cM_{3 2} \cM_{3 4} \cM_{4 1}), \nn
\eea
%%%
\bea
\frac{1}{y_2 y_3} & =& \cM_{1 1} + \expe{4 U} (\cM_{3 3} \cM_{4 1}^2 + \cM_{4 4} \cM_{3 1}^2 - 2 \cM_{3 1} \cM_{3 4} \cM_{4 1}) ,  \\
y_1^{2} & =& \frac{\expe{-4U}}{\cM^2_{3 3}}, \label{y1} \\
%%%
\frac{y_3}{y_2} &=& \cM_{22} - \frac{x_3^2}{y_2 y_3} + \frac{\cM_{23}^2}{\cM_{33}} +\expe{4U} \frac{(\cM_{32} \cM_{34} - \cM_{33} \cM_{42})^2}{\cM_{33}}.
\eea

\subsection*{Details on Weyl reflection}
The truncation to pure five-dimensional Lorentzian gravity corresponds to
taking the six-dimensional metric of the form
\be
ds^2_6 = dy^2 + ds_5^2,
\ee
and setting $\Phi = 0$ and $F_{[3]}=0$.
In terms of the three-dimensional coset scalars, this truncation corresponds to setting
\be
x^I = 0, \qquad y^I = y, \qquad \zeta^I = 0, \qquad \tilde \zeta_I = 0.
\ee
Therefore, the `active' fields are
\be
U, \quad y, \quad \sigma, \quad \zeta^0, \quad \tilde \zeta_0.
\ee
These five fields correspond to an SL(3, $\RR$) truncation of  SO$(4,4)$, generated by the elements
\be
H_0, \quad H_1 + H_2 + H_3, \quad E_{q_{0}}, \quad E_{p^{0}}, \quad E_0, \quad F_{q_{0}}, \quad F_{p^{0}}, \quad F_0. 
\ee  
Under conjugation \eqref{Weyl}, this  SL(3, $\RR$) gets mapped to another  SL(3, $\RR$) generated by,
\be
H_1, \quad H_0 + H_2 + H_3, \quad F_{p^{1}}, \quad E_{p^{0}}, \quad E_1, \quad E_{p^{1}}, \quad F_{p^{0}}, \quad F_1.
\ee  
This new SL(3, $\RR$) corresponds to `active' fields
\be
y^1, \quad  U,  \quad \tilde \zeta_0, \quad \tilde \zeta_1, \quad x^1.
 \label{EuclideanSL3}
\ee

We would like to compare this to a truncation to Euclidean five-dimensional, a metric that arises as
\be
ds^2_6 = -dt^2 + ds_5^2,
\ee
and where the six-dimensional dilaton and the three-form field are set to zero. 
This Euclidean gravity truncation corresponds to setting
\begin{align}
y^1 \ & \ = \ \ f^3 e^{-4U},\\
y^2\ & \ = \ \ y^3 \ \ =  \ \ e^{2U},\\
\ti{\ze}_2 \ & \ = \ \ \ti{\ze}_3 \ \ =0,\\
\ze^0 \ & \ = \ \ \ze^1 \ \ =\ \ \ze^2 \ \ = \ \ \ze^3 \ \ =0, \\
x^2 \ & \ = \ \ x^3 \ \ = \ \ 0, \\
\si \ & \ = \ \ 0,
\end{align}
which conforms to \eqref{EuclideanSL3}.

\subsection*{Three-dimensional seed scalars}
For calculational simplicity we work with coordinate $\varkappa$,
\be
\varkappa := \cos \theta,
\label{xCosTheta}
\ee
 instead of the polar angle $\theta$. For writing equations in the main text we use $\theta$. 
 
We perform a Kaluza-Klein reduction over $y$, $\phi$, and $t$ respectively.
 In three-dimensions we use the convention $\epsilon_{r \varkappa \psi} = + \sqrt{+\det g_\rom{3d}}$. The non-zero scalars in three-dimensions for the metric \eqref{MPinsShifted} are
\begin{align}
& \expe{4U} =  \frac{\tilde \Gamma}{\tilde \Sigma}(1-\varkappa^2),&
& \tilde \zeta_0 = - a_1 a_2 M  \frac{(1-\varkappa^2)^2}{\tilde \Sigma}, \label{scalar1} \\
& \tilde \zeta_1 =  - a_2 M\frac{(1-\varkappa^2)}{\tilde \Sigma},&
& x^1  = -a_1  M \frac{(1-\varkappa^2)}{\tilde \Sigma + M}, \label{scalar2} \\
& y_1 = \frac{\sqrt{\tilde \Sigma \tilde \Gamma}}{\tilde \Sigma + M} \sqrt{1-\varkappa^2},&
& y_2 = y_3 = \sqrt{\frac{\tilde \Gamma}{\tilde \Sigma}} \sqrt{1-\varkappa^2} , \label{scalar3}
\end{align}
where
\bea
\tilde \Sigma  &=& r^2 + a_1^2 (1-\varkappa^2) + a_2^2 \varkappa^2 - M, \label{tSig2}\\
\tilde \Gamma &=& (r^2 + a_2^2) \tilde \Sigma + M a_2^2 (1- \varkappa^2).
\eea
Note that Eq.~\eqref{tSig2} reproduces the relation~\eqref{tSig} introduced earlier.
The three-dimensional base metric is
\be
ds^2_3 = \frac{\tilde \Gamma}{\tilde\Delta} (1-\varkappa^2) dr^2 + \tilde \Gamma d\varkappa^2 + \tilde\Delta \varkappa^2 (1-\varkappa^2) d\psi^2,
\ee
where $\tilde\Delta$ was introduced in \eqref{tDel}.
%$g(r)$ is defined as
%\be
%g(r) =  (r^2 + a_1^2) (r^2 + a_2^2) -M r^2 \equiv   (r^2 -r_+^2)(r^2 - r_-^2).
%\label{gr}
%\ee
%\jvr{This is just $r^2 \tilde\Delta$. Do you think it is necessary to introduce here the function $g(r)$ used by JMaRT?}

\subsection*{Six-dimensional metric} 

Using  scalars \eqref{scalar1}--\eqref{scalar3} we construct the matrix $\cM$. We act on this matrix $\cM$ with the Weyl reflection transformation \eqref{Weyl} and then we perform the charging transformation \eqref{charging}. From the resulting matrix $\cM$ we read all scalars (those obtained in 3d without resorting to dualisation of one-forms) and from the corresponding matrix $\cN$ the three-dimensional one-forms. These pieces allow us to construct the 6d metric. We obtain the over-rotating Cveti\v{c}-Youm metric  \eqref{3charge}. In these calculations 
we have followed the conventions for dimensional reduction and group theory of \cite{Virmani:2012kw}. We have adapted minus signs in the charging transformation  \eqref{charging}, so that the final answer is same as the JMaRT notation. 

A construction of the C-field is more tedious, which we describe next.

%%%%%%%%%%%%%%%%%%%%%%%%%
\section{Construction of the C-field}
\label{app:C2}

In principle all the information about the C-field is also contained in the three-dimensional scalars. Though, in practice, extracting the C-field is tedious. We have proceeded in the following manner.

\subsection*{Overview of the procedure}

An expression for six-dimensional three form $F_{[3]}$ in terms of five-dimensional fields is \cite{Virmani:2012kw},
\be
F_{[3]}^{\rom{(6d)}} = - (h^3)^{-2} \star_5 d A^3_{[1]}  + d A_{[1]}^2 \wedge (d y +  A_{[1]}^1).
\label{F36d}
\ee
In order to compute $F_{[3]}^{\rom{(6d)}}$   we need $(i)$ an explicit expression for the dilatonic scalar $h^3$,  cf.~\eqref{dilatonic_scalars}, $(ii)$ five-dimensional metric to perform the hodge star, and $(iii)$  the three one-forms in five-dimensions.

The dilatonic scalar $h^3$ can be obtained from values of the scalars $y^I$ from  the final matrix $\cM_\rom{final}$. We get
\be
h^3  = \left( \frac{\tilde H_p \tilde H_1}{\tilde H_5^{2}}\right)^\frac{1}{3},
\ee
where
\be
\tilde H_i = r^2 + a_1^2 (1-\varkappa^2) + a_2^2 \varkappa^2 + M s_i^2. 
\ee

\subsection*{Five-dimensional metric} 
The following form of the five-dimensional metric is quite useful \cite{Gimon:2007ps} to perform the Hodge star operation,
\be
ds^2  = - F^2 f (f-M) (d t + k)^2 + F^{-1}ds^2_\rom{base}.
\label{5dmetric}
\ee
It is obtained by dimensional reduction of the 6d dimensional metric  \eqref{3charge} over the $y$-direction.  The four-dimensional base metric in \eqref{5dmetric} is
\bea
ds^2_\rom{base}  &=& \frac{r^2}{(r^2 + a_1^2)(r^2 + a_2^2)-M r^2 } dr^2 + \frac{d\varkappa^2}{1-\varkappa^2}  \nn \\ 
& & + \  (f(f-M))^{-1}\Bigg{\{} (f(f-M) + f (a_2^2 -a_1^2) (1-\varkappa^2) + M a_1^2 (1-\varkappa^2) ) (1-\varkappa^2) d\phi^2 \nn \\  
& & + \ (f(f-M) + f (a_1^2 -a_2^2) \varkappa^2 + M a_2^2  \varkappa^2) \varkappa^2 d\psi^2 \nn \\ 
& &  + \  2 M a_1 a_2 (1-\varkappa^2)\varkappa^2 d \phi d \psi \Bigg{\}}.
\eea
The one form $k$ in \eqref{5dmetric} is
\bea
k = \left[\frac{M s_1 s_5 s_p}{f} a_1  -\frac{M c_1 c_5 c_p}{f- M} a_2  \right] (1-\varkappa^2) d \phi 
+  \left[ \frac{M s_1 s_5 s_p}{f} a_2-\frac{M c_1 c_5 c_p}{f- M} a_1 \right]  \varkappa^2 d \psi,
 \eea
and the functions $F$ and $f$ are,
\bea
F &=& (\tilde H_1 \tilde H_5 \tilde H_p)^{-1/3},\\
f&=& r^2 + a_1^2 (1-\varkappa^2) + a_2^2 \varkappa^2.
\eea

\subsection*{Five-dimensional one forms} 
All three one-forms in five-dimensions are required for the construction of three-form field strength in six-dimensions. These one-forms  (for $I = 1,2,3$), obtained using the matrices $\cM$ and $\cN$, are
\be
A^I =  A^I_\psi d \psi + A^I_t d t + A^I_\phi d \phi,
\ee
where
\begin{align}
 A^1_t &= -\frac{M s_p c_p}{\tilde H_p}, &
 A^2_t &= +\frac{M s_1 c_1}{\tilde H_1}, &
 A^3_t &= -\frac{M s_5 c_5}{\tilde H_5}, &
\end{align}
and
\begin{align}
 A^1_\phi &= \frac{M (a_1 c_p s_1 s_5 - a_2 s_p c_1 c_5)(1-\varkappa^2)}{\tilde H_p} &
 A^1_\psi &= \frac{M (a_2 c_p s_1 s_5 - a_1 s_p c_1 c_5)\varkappa^2}{\tilde H_p}\\
 A^2_\phi &= -\frac{M (a_1 s_p c_1 s_5 - a_2 c_p s_1 c_5)(1-\varkappa^2)}{\tilde H_1} &
 A^2_\psi &= -\frac{M (a_2 s_p c_1 s_5 - a_1 c_p s_1 c_5)\varkappa^2}{\tilde H_1}\\
 A^3_\phi &= \frac{M (a_1 s_p s_1 c_5 - a_2 c_p c_1 s_5) (1-\varkappa^2)}{\tilde H_5} &
 A^3_\psi &= \frac{M (a_2 s_p s_1 c_5 - a_1 c_p c_1 s_5)\varkappa^2}{\tilde H_5}.
\end{align}
Some of our signs are different from those of reference \cite{Gimon:2007ps}, but this is simply because some of our conventions are different\footnote{Note that we use the convention $\epsilon_{r \varkappa \psi} = + \sqrt{+\det g_\rom{3d}}$, where $\varkappa= \cos \theta$.}  and our calculations are organised differently.

\subsection*{Final answer}

Given these expressions it is straightforward, if somewhat tedious, to implement \eqref{F36d}.
We find in six-dimensions $F_{[3]}$ field has 12 independent components. The first six, coming from the first term in \eqref{F36d}, 
$
- (h^3)^{-2} \star_5 d A^3_{[1]},
$  
are
\be
F_{r\phi t}, F_{r\phi \psi}, F_{r t \psi},
F_{\varkappa \phi t}, F_{\varkappa \phi \psi}, F_{\varkappa t \psi},
\ee 
and the next six coming from the second term, $d A_{[1]}^2 \wedge (d y +  A_{[1]}^1)$, are 
\be
F_{r\phi y}, F_{r t y}, F_{r \psi y},
F_{\varkappa \phi y}, F_{\varkappa t y}, F_{\varkappa \psi y}.
\ee

From the resulting F-field a C-field can be constructed by appropriate integrations. An answer is
\be
C_2  = C_{ty} \ d t \wedge d y + C_{t \phi} \ dt \wedge d \phi  + C_{t \psi} \ dt \wedge d \psi  + C_{y \phi} \ d y \wedge d \phi   + C_{\psi \phi} \ d\psi \wedge d \phi + C_{y \psi} \ dy \wedge d \psi, 
\ee
where
\begin{align}
C_{ty}  & =  +\frac{M s_1 c_1}{\tilde H_1}, \label{CFieldI} &
C_{\psi \phi} &=  +\frac{M }{\tilde H_1} s_5 c_5 \left(r^2 + a_2^2 + M s_1^2\right) \varkappa^2, \\
C_{t \psi} &= -\frac{M }{\tilde H_1} \left( a_2 s_5 c_1 c_p - a_1 c_5 s_1 s_p\right) \varkappa^2, &
C_{t \phi} &=- \frac{M }{\tilde H_1} \left( a_1 s_5 c_1 c_p - a_2 c_5 s_1 s_p\right) (1-\varkappa^2), \nn \\
C_{y\psi} &= - \frac{M }{\tilde H_1} \left( a_1 c_5 s_1 c_p - a_2 s_5 c_1 s_p\right) \varkappa^2, &
C_{y \phi} &=  - \frac{M}{\tilde H_1} \left( a_2 c_5 s_1 c_p - a_1 s_5 c_1 s_p\right)(1-\varkappa^2) \nn.
\label{CFieldF}
\end{align}
These expressions  match the corresponding expressions in \cite{Jejjala:2005yu} upto an over-all minus sign (which is convention dependent). In the main text, cf.~\eqref{RR}, we have flipped the over-all minus sign, and have employed the polar angle $\theta$ instead of $\varkappa$.

%%%%%%%%%%%%%%%%%%%%%%%%%%%%%
\section{Rod structure of the Cveti\v{c}-Youm metric}
\label{rod_structure}

Our goal here is to understand the rod structures of the  Cveti\v{c}-Youm metric, in particular the two cases $(i)$ black hole and $(ii)$ fuzzball. 

We recall that solutions of the vacuum Einstein equations in $d$ dimensions with $d-2$ commuting Killing vector fields are classified according to their rod structure: the rods correspond to line sources for a generalised Poisson equation that determines the Killing metric (see appendix~\ref{app:MPins}). In coordinates adapted to the isometries the metric depends explicitly only on two variables, the canonical coordinates $(\rho,z)$, and the rods are located at $\rho=0$. They are physically interpreted as the set of spacetime points where some Killing vector --- the associated rod direction --- degenerates. In particular if the rod is spacelike and extends to $z=\pm\infty$ this indicates an axis of rotation. If the rod is finite and timelike (spacelike) it signals an event horizon (Kaluza-Klein bubble). We refer to \cite{Emparan:2001wk, Harmark:2004rm} for further details.

The above description of rod structures applies only in vacuum, {\it a priori}. Consequently, there is no guarantee that the Cveti\v{c}-Youm solution is amenable to such a treatment when the charges $\delta_p, \delta_1$ and $\delta_5$ are non vanishing. However, we will now see that the rod structure can also be defined for this class of metrics. Since for the JMaRT fuzzball, the $y$ direction shrinks to zero size in the interior of the spacetime,  the analysis of the rod structure is best done in six dimensions. Our starting point is the metric \eqref{3charge}.
For this discussion the order of the Killing coordinates we use is $(t,\phi, \psi, y)$.

\subsection*{Case 1: Black Holes}  
\begin{figure}[t!]
\centering{
\includegraphics[width=9cm]{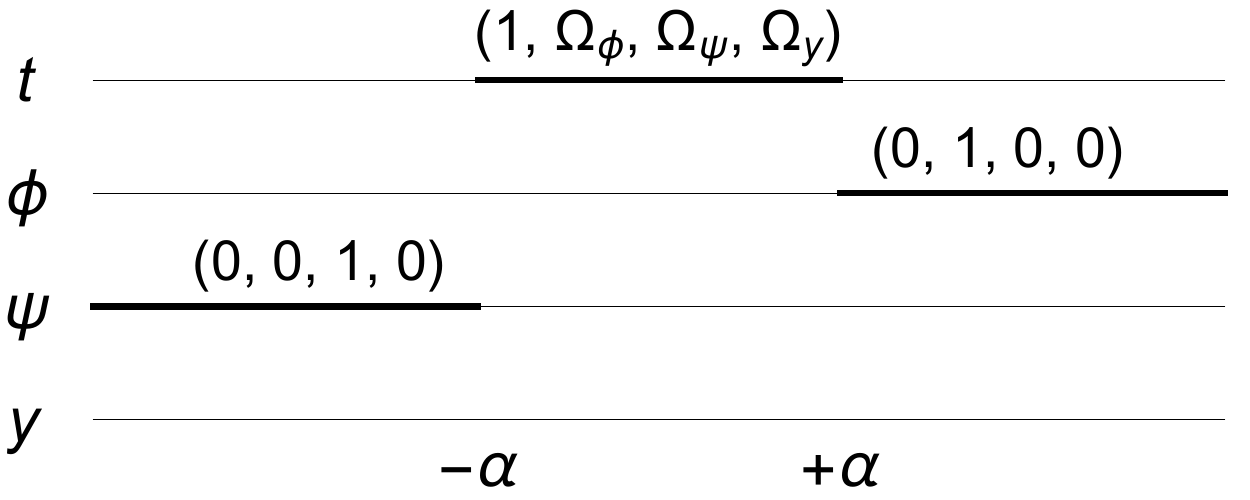}}
\bigskip
\caption{Rod diagram for the Cveti\v{c}-Youm  black hole. The direction for each rod is indicated above the corresponding segment.}
\label{rodBH}
\end{figure}

The Cveti\v{c}-Youm metric describes black holes when $M > (a_1+a_2)^2$.
To analyze the rod structure it is convenient to introduce the prolate spherical coordinates  $(u,v)$ and the canonical coordinates  $(\rho,z)$. In the present  case the coordinate transformation relating the radial coordinates $(r, \theta)$ used in metric \eqref{3charge} to the prolate spherical coordinates $(u,v)$ is 
\bea
r^2 &=&\frac{1}{2}\left(M+ 4u\alpha -a_1^2 -a_2^2 \right), \label{RU}\\ 
\cos^2 \theta &=&  \frac{1}{2}(1-v), \label{ThetaV}
\eea
%\jvr{I flipped the sign in front of $v$, to be consistent with \eqref{prospherical} --- see new comment below. This has no effect on \eqref{EqU} and \eqref{EqV} as long as $z\to-z$ simultaneously. The rod directions also remain unchanged.}
where 
\begin{equation}
\alpha= \frac{1}{4} \sqrt{M-(a_1 +a_2)^2} \sqrt{M-(a_1 -a_2)^2} .
\label{alphaMa1a2}
\end{equation}
We take $a_1 \ge a_2 \ge 0$. Thus $\alpha > 0$.
The canonical coordinates $(\rho,z)$ are related to the prolate coordinates as 
\bea
u &=&\frac{\sqrt{\rho^2 +(z+\alpha)^2}+\sqrt{\rho^2 +(z-\alpha)^2}}{2\alpha}, \label{EqU}\\
v &=&\frac{\sqrt{\rho^2 +(z+\alpha)^2}-\sqrt{\rho^2 +(z-\alpha)^2}}{2\alpha}. \label{EqV}
\eea
Note that Eqs.~(\ref{RU}-\ref{ThetaV}) and~\eqref{alphaMa1a2} above are the inverses of~\eqref{prospherical} and \eqref{Ma1a2alpha}, respectively, upon implementation of the shift transformation~(\ref{shift1}--\ref{shift2}). This makes $r^2\Delta \to r^2\tilde\Delta$, $\cos(2\theta)\to-\cos(2\theta)$ and consequently $(u,v)\to(u,-v)$. This implies $(\rho,z)\to(\rho,-z)$ according to Eqs.~\eqref{canonicalpro}, which are just the inverses of Eqs.~(\ref{EqU}--\ref{EqV}).

 The first rod $\rho = 0, z \in (-\infty, -\alpha)$ corresponds to the degeneration of the $\psi$ circle at $\theta =\pi/2$, i.e.,  its  rod vector is $(0,0,1,0)$. 
 The second rod $\rho = 0, z \in (-\alpha, \alpha)$ corresponds to  the horizon with rod vector $(1, \Omega_\phi, \Omega_\psi,  \Omega_y)$. The Killing vector that degenerates at the horizon is \be
 \xi= \frac{\partial}{\partial t} + \Omega_\phi \frac{\partial}{\partial \phi} + \Omega_\psi \frac{\partial}{\partial \psi}+ \Omega_y \frac{\partial}{\partial y}.
 \ee
Explicit expressions for $\Omega_\phi$, $\Omega_\psi$, and $\Omega_y$ are (see also \cite{Cvetic:1997uw}),
\begin{eqnarray}
\Omega_\phi 
&=& +\frac{1}{\gamma} \left[\frac{a_1 -a_2}{\sqrt{M-(a_1 -a_2)^2}}- \frac{a_1 +a_2}{\sqrt{M-(a_1 +a_2)^2}}\right], \\
\Omega_\psi 
&=& -\frac{1}{\gamma} \left[\frac{a_1 -a_2}{\sqrt{M-(a_1 -a_2)^2}}+\frac{a_1 +a_2}{\sqrt{M-(a_1 +a_2)^2}}\right],  \\
\Omega_y &=& \frac{M }{\gamma} \left[\frac{c_1 c_5 s_p - s_1 s_5 c_p}{\sqrt{M-(a_1 -a_2)^2}}+\frac{c_1 c_5 s_p + s_1 s_5 c_p}{\sqrt{M-(a_1 +a_2)^2}} \right],  \nn
\end{eqnarray} 
where
\be
\gamma =  M \left[ \frac{ c_1 c_5 c_p-s_1 s_5 s_p }{\sqrt{M-(a_1 -a_2)^2}} + \frac{ c_1 c_5 c_p+s_1 s_5 s_p }{\sqrt{M-(a_1 +a_2)^2}} \right].
\ee
 The third rod $\rho=0, z \in (\alpha, \infty)$ corresponds to the degeneration of the $\phi $ circle at $\theta =0$, i.e.,  its rod vector is $(0,1,0,0)$.  
The rod diagram is shown in figure \ref{rodBH}.

\subsection*{Case 2: Fuzzballs}

\begin{figure}[t!]
\centering{
\includegraphics[width=9cm]{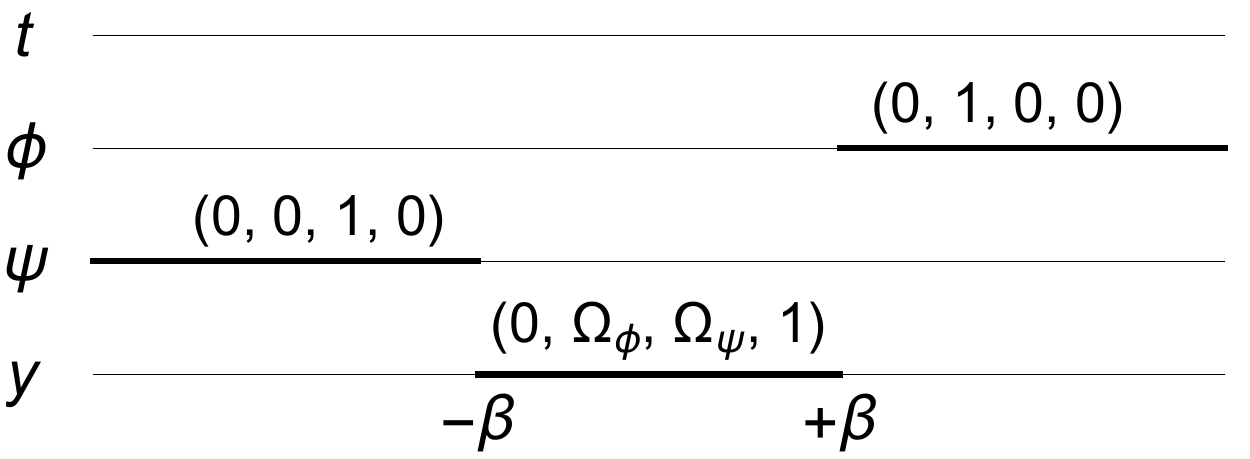}}
\bigskip
\caption{Rod diagram for the JMaRT fuzzball. The direction for each rod is indicated above the corresponding segment.}
\label{rodFuzzball}
\end{figure}

For the smooth solitonic fuzzball solutions we have $(a_1 -a_2)^2 > M$. 
The end points  of the rod on the $z$-axis are at $\pm\beta$ where 
\be
\beta =  \frac{1}{4} \sqrt{(a_1 + a_2)^2-M} \sqrt{(a_1 - a_2)^2-M}.
\ee 
Note that $\beta > 0$. We introduce the prolate and the canonical coordinates exactly in the same manner as in the black hole case. The radial coordinates $(r, \theta)$ used in metric \eqref{3charge} are related to the prolate spherical coordinates $(u,v)$ via
\bea
r^2 &=&\frac{1}{2}\left(M+ 4u\beta -a_1^2 -a_2^2 \right), \\ 
\cos^2 \theta &=&  \frac{1}{2}(1-v),
\eea
%\jvr{I flipped the sign in front of $v$, as in the black hole case.}
and the canonical coordinates $(\rho,z)$ are related to the prolate coordinates as 
\bea
u &=&\frac{\sqrt{\rho^2 +(z+\beta)^2}+\sqrt{\rho^2 +(z-\beta)^2}}{2\beta}, \\
v &=&\frac{\sqrt{\rho^2 +(z+\beta)^2}-\sqrt{\rho^2 +(z-\beta)^2}}{2\beta}.
\eea

 As in the black hole case, the first rod $z \in (-\infty, -\beta)$ corresponds to the degeneration of the $\psi$ circle at $\theta =\pi/2$, i.e.,  its  rod vector is $(0,0,1,0)$.  The third rod $z \in (\beta, \infty)$ corresponds the degeneration of the $\phi $ circle at $\theta =0$, i.e.,  its rod vector is $(0,1,0,0)$.   The second rod $\rho = 0, z \in (-\beta, \beta)$ corresponds to the degeneration of the $y$ direction.
The determinant of the $(4\times4)$ Killing matrix over coordinates $(t,\phi, \psi, y)$ vanishes at $\rho=0$, which in terms of the original radial coordinate translates into 
\be
 r^2=r_+^2:= \frac{M+ 4 \beta -a_1^2-a_2^2}{2}.
 \label{rplus}
\ee

The fuzzball construction \cite{Jejjala:2005yu} further requires that the determinant of the $(3 \times 3)$ Killing matrix  over purely spatial directions $(\phi, \psi, y)$ vanishes at $\rho=0,  z \in (-\beta, \beta)$, i.e., at $r=r_+$. So, we consider $t = \mbox{const}$ slice along with $r=r_+$. The determinant of the $(3 \times 3)$ Killing matrix vanishes for
\be
M=a_1^2+ a_2^2- a_1 a_2 \frac{(s_1^2  s_5^2 s_p^2+ c_1^2 c_5^2 c_p^2)}{s_1 s_5 s_p c_1 c_5 c_p}.
\ee 
Substituting this value of $M$ in \eqref{rplus} we get,
\begin{equation}
r_+^2=-a_1 a_2 \frac{s_1 s_5 s_p}{c_1 c_5 c_p}.
\end{equation}
The Killing vector that degenerates at the second rod $\rho = 0, z \in (-\beta, \beta)$ is
\be
\xi= \frac{\partial}{\partial y} + \Omega_\phi \frac{\partial}{\partial \phi} + \Omega_\psi \frac{\partial}{\partial \psi},
\ee
 with 
 \begin{align}
 \Omega_\phi &=\frac{s_p c_p}{ a_2 c_1 c_5 c_p-a_1 s_1 s_5 s_p}, &
  \Omega_\psi &= \frac{s_p c_p}{ a_1 c_1 c_5 c_p-a_2 s_1 s_5 s_p}. 
 \end{align}
The rod diagram is shown in figure \ref{rodFuzzball}.

%%%%%%%%%%%%%%%%%%%%%%%%%%%
%%%%%%%%%%%%%%%%%%%%%%%%%%%


\begin{thebibliography}{99}
  
  %\cite{Giusto:2004kj}
\bibitem{Giusto:2004kj} 
  S.~Giusto and S.~D.~Mathur,
  ``Geometry of D1-D5-P bound states,''
  Nucl.\ Phys.\ B {\bf 729}, 203 (2005)
  [hep-th/0409067].
  %%CITATION = HEP-TH/0409067;%%
  %69 citations counted in INSPIRE as of 09 Nov 2015
  
  %\cite{Bena:2005va}\cite{Berglund:2005vb}
\bibitem{Bena:2005va} 
  I.~Bena and N.~P.~Warner,
``Bubbling supertubes and foaming black holes,''
  Phys.\ Rev.\ D {\bf 74}, 066001 (2006)
  [hep-th/0505166].
  %%CITATION = HEP-TH/0505166;%%
  %168 citations counted in INSPIRE as of 09 Nov 2015

%\cite{Berglund:2005vb}
\bibitem{Berglund:2005vb} 
  P.~Berglund, E.~G.~Gimon and T.~S.~Levi,
``Supergravity microstates for BPS black holes and black rings,''
  JHEP {\bf 0606}, 007 (2006)
  [hep-th/0505167].
  %%CITATION = HEP-TH/0505167;%%
  %160 citations counted in INSPIRE as of 09 Nov 2015
  
  
  %\cite{Jejjala:2005yu}
\bibitem{Jejjala:2005yu} 
  V.~Jejjala, O.~Madden, S.~F.~Ross and G.~Titchener,
  ``Non-supersymmetric smooth geometries and D1-D5-P bound states,''
  Phys.\ Rev.\ D {\bf 71}, 124030 (2005)
  doi:10.1103/PhysRevD.71.124030
  [hep-th/0504181].
  %%CITATION = doi:10.1103/PhysRevD.71.124030;%%
  %113 citations counted in INSPIRE as of 05 févr. 2016
  
  %\cite{Giusto:2007tt}
\bibitem{Giusto:2007tt} 
  S.~Giusto, S.~F.~Ross and A.~Saxena,
  ``Non-supersymmetric microstates of the D1-D5-KK system,''
  JHEP {\bf 0712}, 065 (2007)
  doi:10.1088/1126-6708/2007/12/065
  [arXiv:0708.3845 [hep-th]].
  %%CITATION = doi:10.1088/1126-6708/2007/12/065;%%
  %51 citations counted in INSPIRE as of 04 févr. 2016
  
  %\cite{AlAlawi:2009qe}
\bibitem{AlAlawi:2009qe} 
  J.~H.~Al-Alawi and S.~F.~Ross,
  ``Spectral Flow of the Non-Supersymmetric Microstates of the D1-D5-KK System,''
  JHEP {\bf 0910}, 082 (2009)
  doi:10.1088/1126-6708/2009/10/082
  [arXiv:0908.0417 [hep-th]].
  %%CITATION = doi:10.1088/1126-6708/2009/10/082;%%
  %19 citations counted in INSPIRE as of 05 févr. 2016

%\cite{Banerjee:2014hza}
\bibitem{Banerjee:2014hza} 
  S.~Banerjee, B.~D.~Chowdhury, B.~Vercnocke and A.~Virmani,
  ``Non-supersymmetric Microstates of the MSW System,''
  JHEP {\bf 1405}, 011 (2014)
  doi:10.1007/JHEP05(2014)011
  [arXiv:1402.4212 [hep-th]].
  %%CITATION = doi:10.1007/JHEP05(2014)011;%%
  %9 citations counted in INSPIRE as of 05 Feb 2016
  
  %\cite{Bena:2009qv}
\bibitem{Bena:2009qv} 
  I.~Bena, S.~Giusto, C.~Ruef and N.~P.~Warner,
  ``A (Running) Bolt for New Reasons,''
  JHEP {\bf 0911}, 089 (2009)
  doi:10.1088/1126-6708/2009/11/089
  [arXiv:0909.2559 [hep-th]].
  %%CITATION = doi:10.1088/1126-6708/2009/11/089;%%
  %47 citations counted in INSPIRE as of 05 févr. 2016
  
  %\cite{Bossard:2014yta}
\bibitem{Bossard:2014yta} 
  G.~Bossard and S.~Katmadas,
  ``A bubbling bolt,''
  JHEP {\bf 1407}, 118 (2014)
  doi:10.1007/JHEP07(2014)118
  [arXiv:1405.4325 [hep-th]].
  %%CITATION = doi:10.1007/JHEP07(2014)118;%%
  %7 citations counted in INSPIRE as of 05 févr. 2016
  
  %\cite{Bena:2015drs}
\bibitem{Bena:2015drs} 
  I.~Bena, G.~Bossard, S.~Katmadas and D.~Turton,
  ``Non-BPS multi-bubble microstate geometries,''
  arXiv:1511.03669 [hep-th].
  %%CITATION = ARXIV:1511.03669;%%
  %2 citations counted in INSPIRE as of 05 févr. 2016


%\cite{Gimon:2007ps}
\bibitem{Gimon:2007ps} 
  E.~G.~Gimon, T.~S.~Levi and S.~F.~Ross,
  ``Geometry of non-supersymmetric three-charge bound states,''
  JHEP {\bf 0708}, 055 (2007)
  [arXiv:0705.1238 [hep-th]].
  %%CITATION = ARXIV:0705.1238;%%
  %24 citations counted in INSPIRE as of 09 Nov 2015
  
  
  %\cite{Chen:2010zu}
\bibitem{Chen:2010zu} 
  Y.~Chen and E.~Teo,
  ``Rod-structure classification of gravitational instantons with U(1) $\times$ U(1) isometry,''
  Nucl.\ Phys.\ B {\bf 838}, 207 (2010)
  [arXiv:1004.2750 [gr-qc]].
  %%CITATION = ARXIV:1004.2750;%%
  %25 citations counted in INSPIRE as of 10 Nov 2015




\bibitem{Emparan:2001wk} 
R.~Emparan, H.~S.~Reall,
``Generalized Weyl solutions,''
  Phys.\ Rev.\  {\bf D65}, 084025 (2002).
  [hep-th/0110258].


%\cite{Harmark:2004rm}
\bibitem{Harmark:2004rm} 
  T.~Harmark,
  ``Stationary and axisymmetric solutions of higher-dimensional general relativity,''
  Phys.\ Rev.\ D {\bf 70}, 124002 (2004)
%  doi:10.1103/PhysRevD.70.124002
  [hep-th/0408141].
  %%CITATION = doi:10.1103/PhysRevD.70.124002;%%
  %155 citations counted in INSPIRE as of 10 févr. 201


%\cite{Hollands:2007aj}
\bibitem{Hollands:2007aj} 
  S.~Hollands and S.~Yazadjiev,
  ``Uniqueness theorem for 5-dimensional black holes with two axial Killing fields,''
  Commun.\ Math.\ Phys.\  {\bf 283}, 749 (2008)
  doi:10.1007/s00220-008-0516-3
  [arXiv:0707.2775 [gr-qc]].
  %%CITATION = doi:10.1007/s00220-008-0516-3;%%
  %111 citations counted in INSPIRE as of 10 Feb 2016
  
  %\cite{Bobev:2009kn}
\bibitem{Bobev:2009kn} 
  N.~Bobev and C.~Ruef,
  ``The Nuts and Bolts of Einstein-Maxwell Solutions,''
  JHEP {\bf 1001}, 124 (2010)
  [arXiv:0912.0010 [hep-th]].
  %%CITATION = ARXIV:0912.0010;%%
  %34 citations counted in INSPIRE as of 10 Nov 2015
  
  %\cite{Bobev:2009kn}\cite{Bossard:2014ola}
\bibitem{Bossard:2014ola} 
  G.~Bossard and S.~Katmadas,
  ``Floating JMaRT,''
  JHEP {\bf 1504}, 067 (2015)
  [arXiv:1412.5217 [hep-th]].
  %%CITATION = ARXIV:1412.5217;%%
  %3 citations counted in INSPIRE as of 10 Nov 2015


%\cite{Katsimpouri:2014ara}
\bibitem{Katsimpouri:2014ara} 
  D.~Katsimpouri, A.~Kleinschmidt and A.~Virmani,
  ``An Inverse Scattering Construction of the JMaRT Fuzzball,''
  JHEP {\bf 1412}, 070 (2014)
  %doi:10.1007/JHEP12(2014)070
  [arXiv:1409.6471 [hep-th]].
  
 

\bibitem{BZ}
V. A. Belinski and V. E. Zakharov, ``Integration Of The Einstein Equations By The Inverse
Scattering Problem Technique And The Calculation Of The Exact Soliton Solutions,'' Sov.
Phys. JETP {\bf 48}, 985 (1978) [Zh. Eksp. Teor. Fiz. {\bf 75}, 1953 (1978)].  \\  \vskip -0.5cm
V. A. Belinski, V. E. Sakharov, ``Stationary Gravitational Solitons with Axial Symmetry,''
 Sov. Phys. JETP {\bf 50} 1 ,1979, [Zh.Eksp.Teor.Fiz. {\bf 77} 3,1979].   \\  \vskip -0.5cm V. A. Belinski and E. Verdaguer, \emph{ Gravitational solitons}, Cambridge University Press, 2001.

%\cite{Pomeransky:2005sj}
\bibitem{Pomeransky:2005sj} 
  A.~A.~Pomeransky,
  ``Complete integrability of higher-dimensional Einstein equations with additional symmetry, and rotating black holes,''
  Phys.\ Rev.\ D {\bf 73}, 044004 (2006)
  %doi:10.1103/PhysRevD.73.044004
  [hep-th/0507250].
  
  %\cite{Breitenlohner:1986um}
\bibitem{Breitenlohner:1986um} 
  P.~Breitenlohner and D.~Maison,
 ``On the Geroch Group,''
  Annales Poincare Phys.\ Theor.\  {\bf 46}, 215 (1987).
  %%CITATION = AHPAA,46,215;%%
  %99 citations counted in INSPIRE as of 05 févr. 2016
\bibitem{Katsimpouri:2013wka} 
  D.~Katsimpouri, A.~Kleinschmidt and A.~Virmani,
  ``An inverse scattering formalism for STU supergravity,''
  JHEP {\bf 1403}, 101 (2014)
  doi:10.1007/JHEP03(2014)101
  [arXiv:1311.7018 [hep-th]].
  %%CITATION = doi:10.1007/JHEP03(2014)101;%%
  %5 citations counted in INSPIRE as of 05 Feb 2016
  %\cite{Katsimpouri:2012ky}
\bibitem{Katsimpouri:2012ky} 
  D.~Katsimpouri, A.~Kleinschmidt and A.~Virmani,
  ``Inverse Scattering and the Geroch Group,''
  JHEP {\bf 1302}, 011 (2013)
  doi:10.1007/JHEP02(2013)011
  [arXiv:1211.3044 [hep-th]].
  %%CITATION = doi:10.1007/JHEP02(2013)011;%%
  %8 citations counted in INSPIRE as of 05 Feb 2016
  
  %\cite{Chakrabarty:2014ora}
\bibitem{Chakrabarty:2014ora} 
  B.~Chakrabarty and A.~Virmani,
  ``Geroch Group Description of Black Holes,''
  JHEP {\bf 1411}, 068 (2014)
  doi:10.1007/JHEP11(2014)068
  [arXiv:1408.0875 [hep-th]].
  %%CITATION = doi:10.1007/JHEP11(2014)068;%%
%\cite{Katsimpouri:2013wka}


%\cite{Cvetic:1996xz}
\bibitem{Cvetic:1996xz} 
  M.~Cveti\v{c} and D.~Youm,
  %``General rotating five-dimensional black holes of toroidally compactified heterotic string,''
  Nucl.\ Phys.\ B {\bf 476}, 118 (1996)
  doi:10.1016/0550-3213(96)00355-0
  [hep-th/9603100].
  %%CITATION = doi:10.1016/0550-3213(96)00355-0;%%
  %257 citations counted in INSPIRE as of 25 Feb 2016



%\cite{Youm:1997hw}
\bibitem{Youm:1997hw} 
  D.~Youm,
 ``Black holes and solitons in string theory,''
  Phys.\ Rept.\  {\bf 316}, 1 (1999)
  doi:10.1016/S0370-1573(99)00037-X
  [hep-th/9710046].
  %%CITATION = doi:10.1016/S0370-1573(99)00037-X;%%
  %184 citations counted in INSPIRE as of 22 Mar 2016

%\cite{Bossard:2009we}
\bibitem{Bossard:2009we} 
  G.~Bossard, Y.~Michel and B.~Pioline,
  %``Extremal black holes, nilpotent orbits and the true fake superpotential,''
  JHEP {\bf 1001}, 038 (2010)
  doi:10.1007/JHEP01(2010)038
  [arXiv:0908.1742 [hep-th]].
  
  %\cite{Virmani:2012kw}
\bibitem{Virmani:2012kw} 
  A.~Virmani,
  ``Subtracted Geometry From Harrison Transformations,''
  JHEP {\bf 1207}, 086 (2012)
  [arXiv:1203.5088 [hep-th]].
  %%CITATION = ARXIV:1203.5088;%%
  %27 citations counted in INSPIRE as of 10 Nov 2015




%\cite{Izumi:2007qx}
\bibitem{Izumi:2007qx} 
  K.~Izumi,
  ``Orthogonal black di-ring solution,''
  Prog.\ Theor.\ Phys.\  {\bf 119}, 757 (2008)
  %doi:10.1143/PTP.119.757
  [arXiv:0712.0902 [hep-th]].

%\cite{Emparan:2008eg}
\bibitem{Emparan:2008eg} 
  R.~Emparan and H.~S.~Reall,
  ``Black Holes in Higher Dimensions,''
  Living Rev.\ Rel.\  {\bf 11}, 6 (2008)
  %doi:10.12942/lrr-2008-6
  [arXiv:0801.3471 [hep-th]].

%\cite{Elvang:2007rd}
\bibitem{Elvang:2007rd}
  H.~Elvang and P.~Figueras,
  ``Black Saturn,''
  JHEP {\bf 0705} (2007) 050
  %doi:10.1088/1126-6708/2007/05/050
  [hep-th/0701035].

%\cite{Rocha:2013qya}
\bibitem{Rocha:2013qya} 
  J.~V.~Rocha, M.~J.~Rodriguez, O.~Varela and A.~Virmani,
  ``Charged black rings from inverse scattering,''
  Gen.\ Rel.\ Grav.\  {\bf 45}, 2099 (2013)
  doi:10.1007/s10714-013-1586-x
  [arXiv:1305.4969 [hep-th]].
  %%CITATION = doi:10.1007/s10714-013-1586-x;%%
  %4 citations counted in INSPIRE as of 04 févr. 2016
  
  
  %\cite{Sahay:2013xda}
\bibitem{Sahay:2013xda} 
  A.~Sahay and A.~Virmani,
  ``Subtracted Geometry from Harrison Transformations: II,''
  JHEP {\bf 1307}, 089 (2013)
  doi:10.1007/JHEP07(2013)089
  [arXiv:1305.2800 [hep-th]].
  %%CITATION = doi:10.1007/JHEP07(2013)089;%%
  %4 citations counted in INSPIRE as of 11 févr. 2016
  
  %\cite{Chow:2014cca}
\bibitem{Chow:2014cca} 
  D.~D.~K.~Chow and G.~Comp\`ere,
  ``Black holes in N=8 supergravity from SO(4,4) hidden symmetries,''
  Phys.\ Rev.\ D {\bf 90}, no. 2, 025029 (2014)
  [arXiv:1404.2602 [hep-th]].
  %%CITATION = ARXIV:1404.2602;%%
  %11 citations counted in INSPIRE as of 09 Nov 2015
  
%\cite{Gal'tsov:2008sh}
\bibitem{Gal'tsov:2008sh} 
  D.~V.~Gal'tsov and N.~G.~Scherbluk,
  ``Improved generating technique for D=5 supergravities and squashed Kaluza-Klein Black Holes,''
  Phys.\ Rev.\ D {\bf 79}, 064020 (2009)
  [arXiv:0812.2336 [hep-th]].
  %%CITATION = ARXIV:0812.2336;%%
  %29 citations counted in INSPIRE as of 09 Nov 2015


%\cite{Compere:2009zh}
\bibitem{Compere:2009zh} 
  G.~Comp\`ere, S.~de Buyl, E.~Jamsin and A.~Virmani,
  ``G2 Dualities in D=5 Supergravity and Black Strings,''
  Class.\ Quant.\ Grav.\  {\bf 26}, 125016 (2009)
  [arXiv:0903.1645 [hep-th]].
  %%CITATION = ARXIV:0903.1645;%%
  %34 citations counted in INSPIRE as of 09 Nov 2015
  
%\cite{Cvetic:1997uw}
\bibitem{Cvetic:1997uw} 
  M.~Cveti\v{c} and F.~Larsen,
  ``General rotating black holes in string theory: Grey body factors and event horizons,''
  Phys.\ Rev.\ D {\bf 56}, 4994 (1997)
  [hep-th/9705192].
  %%CITATION = HEP-TH/9705192;%%
  %153 citations counted in INSPIRE as of 12 Nov 2015

\end{thebibliography}
\end{document}